\documentclass[aps,prd,preprint,superscriptaddress,showpacs,floatfix,preprintnumbers,nofootinbib]{revtex4-1}

\usepackage[utf8]{inputenc}

\usepackage{cancel}

\usepackage[
    letterpaper, 
    left = 2.5cm, 
    right = 2.5cm,
    top = 2.4cm, 
    bottom = 2.4cm, 
    headsep = 0.5cm, 
    footskip = 1.5cm]{geometry}

\usepackage[dvipsnames]{xcolor}      
\definecolor{lcolor}{rgb}{0.5,0,0}
\definecolor{citcolor}{rgb}{0,0.0,1}

\usepackage[breaklinks,colorlinks,urlcolor=blue,citecolor=citcolor,linkcolor=lcolor,linktoc=all]{hyperref}
\usepackage{color}
\usepackage{graphicx}	
\graphicspath{{./figures/}}
\usepackage[utf8]{inputenc}
\usepackage{amsmath} 
\usepackage{amssymb}
\usepackage[ragged]{footmisc}
\usepackage{slashed} 
\usepackage{float}

\makeatletter
\g@addto@macro\bfseries{\boldmath}
\makeatother

\usepackage{tikz}
\usepackage[customcolors]{hf-tikz}
\usepackage{mciteplus}

\usetikzlibrary{arrows,cd,shapes,decorations.pathmorphing,decorations.markings,shadings}
\tikzset{
  big arrow/.style={
    decoration={markings,mark=at position 1 with {\arrow[scale=4,#1]{>}}},
    postaction={decorate},
    shorten >=0.4pt},
  big arrow/.default=blue}

\newcommand{\Kc}{K}
\newcommand{\Pc}{P}
\newcommand{\Qc}{Q}

\newcommand{\bbone}{\text{\usefont{U}{bbold}{m}{n}1}}
\MakeRobust{\bbone}
\newcommand{\eq}{Eq.~}
\newcommand{\eqs}{Eqs.~}
\newcommand{\Sec}{Sec.~}
\newcommand{\fig}{Fig.~}
\newcommand{\figs}{Figs.~}
\newcommand{\app}{Appendix~}

\newcommand{\nn}{\nonumber \\ }
\newcommand{\nr}[1]{(\ref{#1})}
\renewcommand{\Ref}{Ref.~}
\newcommand{\Refs}{Refs.~}
\newcommand{\Lh}{\Lambda_{\text{h}}}
\newcommand{\Lbar}{\overline{\Lambda}}
\newcommand{\mE}{m_{\rm E}}
\newcommand{\mD}{m_{\rm D}}

\newcommand{\prjoT}{\mathcal{P}_{\text{T}}}
\newcommand{\prjoL}{\mathcal{P}_{\text{L}}}
\newcommand{\prjof}{\mathcal{P}_{D}}

\newcommand{\kt}{{\vec{k}}}
\newcommand{\pt}{{\vec{p}}}

\newcommand{\zerot}{{\vec{0}}}
\renewcommand{\vec}{\mathbf}

\renewcommand{\epsilon}{\varepsilon}

\newcommand{\ud}{\mathrm{d}}

\renewcommand{\Im}{\operatorname{Im}}
\newcommand{\gamE}{\gamma_{\text{E}}}

\renewcommand{\Re}{\operatorname{Re}}
\usepackage[acronym]{glossaries}
\newacronym{LO}{LO}{leading-order}

\newacronym{QM}{QM}{quark matter}

\newacronym{QCD}{QCD}{quantum chromodynamics}

\newacronym{HTL}{HTL}{Hard-Thermal-Loop}

\newacronym{UV}{UV}{ultraviolet}

\newacronym{IR}{IR}{infrared}

\newcommand{\che}{e}
\newcommand{\exe}{\mathrm{e}}                                   

\newcommand{\nocontentsline}[3]{}
\newcommand{\tocless}[2]{\bgroup\let\addcontentsline=\nocontentsline#1{#2}\egroup}

\renewcommand{\log}{\ln}

\begin{document}

\title{Soft photon propagation in a hot and dense medium to next-to-leading order}

\preprint{HIP-2022-8/TH}
\author{Tyler Gorda}
\affiliation{Technische Universit\"{a}t Darmstadt, Department of Physics, 64289 Darmstadt, Germany}
\affiliation{ExtreMe Matter Institute EMMI and Helmholtz Research Academy for FAIR, GSI Helmholtzzentrum f\"ur Schwerionenforschung GmbH, 64291 Darmstadt, Germany}
\author{Aleksi Kurkela}
\affiliation{Faculty of Science and Technology, University of Stavanger, 4036 Stavanger, Norway}
\author{Juuso Österman}
\affiliation{Department of Physics and Helsinki Institute of Physics,
P.O.~Box 64, FI-00014 University of Helsinki, Finland}
\author{Risto Paatelainen}
\affiliation{Department of Physics and Helsinki Institute of Physics,
P.O.~Box 64, FI-00014 University of Helsinki, Finland}
\author{Saga Säppi}
\affiliation{European Centre for Theoretical Studies in Nuclear Physics and Related Areas (ECT*) and Fondazione Bruno Kessler, Strada delle Tabarelle 286, I-38123, Villazzano (TN), Italy}
\author{Philipp Schicho}
\affiliation{Department of Physics and Helsinki Institute of Physics,
P.O.~Box 64, FI-00014 University of Helsinki, Finland}
\author{Kaapo Seppänen}
\affiliation{Department of Physics and Helsinki Institute of Physics,
P.O.~Box 64, FI-00014 University of Helsinki, Finland}
\author{Aleksi Vuorinen}
\affiliation{Department of Physics and Helsinki Institute of Physics,
P.O.~Box 64, FI-00014 University of Helsinki, Finland}

\begin{abstract}
\noindent
We present the first complete calculation of the soft photon self-energy to the next-to-leading order in a hot and/or dense ultrarelativistic plasma in Quantum Electrodynamics (QED). The calculation is performed within the real-time formalism utilizing dimensional regularization in $4-2\epsilon$ dimensions, while the result is reported including explicit $O(\epsilon)$ terms in the zero-temperature limit. This information is required to extend the perturbative calculation of the pressure of cold and dense QED matter to partial next-to-next-to-next-to-leading order in a weak-coupling expansion, reported in a companion paper. These results pave the way for a  similar future calculation in Quantum Chromodynamics.
\end{abstract}

\maketitle

\tableofcontents

\newpage

\section{Introduction }
\label{sec:intro}
\noindent
Perturbative thermal field theory is a frequently used technique in a number of subfields of theoretical high energy physics, ranging from studies of early Universe cosmology to heavy-ion phenomenology and the physics of neutron stars. In high-order computations, a common and at times problematic issue has to do with the infrared (IR) sensitivity of different physical quantities, originating from the contributions of long-range massless (bosonic) fields and leading to uncancelled divergences in naive perturbation theory (see e.g.~\cite{Ghiglieri:2020dpq} for a review). This has lead to the need to develop both resummation techniques and effective-field-theory methods for taming this so-called ``soft sector'' of various quantum field theories, culminating in the development and application of dimensionally reduced \cite{Appelquist:1981vg,Kajantie:1995dw,Braaten:1995cm} and Hard-Thermal-Loop (HTL) effective theories \cite{Braaten:1989mz,Braaten:1991gm}. The dimensionally reduced effective theories are applicable to static observables at high temperatures $T$ (see e.g.~\cite{Kurkela:2016was}), while the HTL framework is more versatile, remaining functional even at zero temperature and for time-dependent quantities. 

In the context of Quantum Chromodynamics (QCD) at nonzero temperature and/or density, efforts to determine the pressure, or equation of state, have reached such a high order \cite{Kajantie:2002wa,Kajantie:2003ax,Vuorinen:2003fs,DiRenzo:2006nh,Gorda:2021kme,Gorda:2021znl} that physical contributions can no longer be classified as being purely hard (corresponding to scales such as $\pi T$ or $\mu$,
where $\mu$ is the chemical potential of quarks) or soft (scales proportional to electric or magnetic screening masses). Instead, these modes also interact with each other in ways that have recently been characterized at zero temperature \cite{Gorda:2021kme}, leading to the generation of so-called `mixed' contributions to physical quantities. In the particular case of the $T=0$ pressure of cold quark matter, such contributions enter at $O(\alpha_s^3)$ in the strong coupling constant $\alpha_s$, and require the dressing of gluon propagators with self-energies that go beyond the usual leading-order (LO) one-loop HTL expression, encountered already in lower-order computations. 

Given that the LO HTL approximation amounts to studying one-loop self-energies (and vertex functions) in the limit of soft external momenta, there are two independent ways in which one can proceed beyond this limit to next-to-leading order (NLO). On the one hand, we may expand the one-loop self-energy of the full theory beyond LO in powers of the soft momentum. On the other hand, we may consider two-loop corrections at LO in the momentum expansion\footnote{In principle, there is a third possible source of NLO corrections as well, corresponding to soft loop momenta in the one-loop self-energy diagrams. These corrections will, however, turn out to be of subleading order for the calculation performed in this paper~\cite{Mirza:2013ula}.}.
In the context of the Abelian theory Quantum Electrodynamics (QED) and QCD, the one-loop photon and gluon HTL self-energies%
\footnote{The computation was also done including the full kinematics.}
at finite temperature and nonzero density were first computed by Toimela in \Ref\cite{Toimela:1984xy}. Recently, the photon self-energy was extended to two loops at nonzero  temperature but vanishing density in the limit of soft external momenta%
\footnote{In the full kinematics, the computation was performed in \Ref\cite{Jackson:2019mop}.}
and massless fermions~\cite{Carignano:2019ofj}. The main goal of the present paper is to generalize this two-loop computation to nonzero chemical potentials. 

The reason to consider QED in the present work is twofold. First, owing to the lack of self-interactions between photons, the evaluation of the photon two-point function to NLO both in a loop and small-momentum expansion is considerably more straightforward than in QCD. Thus, it makes sense to begin by considering this more tractable --- and yet very nontrivial --- limit, and return to the subtleties related to the non-Abelian nature of QCD later. Second, QED is of course an interesting physical theory on its own, although due to the small value of the fine-structure constant $\alpha_e$, weak-coupling expansions performed there tend to converge considerably better than in QCD. Evaluating the two-loop self-energy in QED enables us to analyze the function at various coupling strengths, and to study the interesting physics contained therein. Furthermore, this allows the evaluation of the mixed contributions to the QED pressure at next-to-next-to-next-to-leading order (N$^3$LO), including the determination of the complete $O(\alpha_e^3 \ln \alpha_e^{ })$-term at zero temperature. This calculation will be described in the associated companion paper~\cite{Seppanen}. 

The present article is organized as follows. In \Sec\ref{sec:organization}, we introduce to the reader both our notation and the technical background of our computation. This includes a detailed description of how the two-point function is evaluated in the real-time formalism of thermal field theory, and an introduction of the general structure of the photon self-energy tensor and its low-momentum HTL expansion. In \Sec\ref{sec:detailedcomputation}, we then present the detailed computation of the one- and two-loop photon self-energies with arbitrary soft external momenta within the framework of dimensional regularization in $4-2\varepsilon$ dimensions. \Sec\ref{sec:results}  on the other hand contains a summary of the main results for the photon self-energies and an analysis of the resulting dispersion relations for transverse and longitudinal photon modes. In \Sec\ref{coldstuff}, we then compute the explicit $O(\varepsilon)$ term for the one- and two-loop self-energies in the limit of soft external momenta at zero temperature. These results become useful when considering higher-order vacuum diagrams for the pressure in which the self-energy appears.
Finally, in \Sec\ref{sec:discussion} we draw our conclusions and consider the implications of our work, while many technical details of the calculations are explained in the Appendices. 

\section{Organizing the computation}
\label{sec:organization}
\noindent
In this section, we provide the reader with the background toolkit needed to follow the details of our computation. In particular, we introduce the formalism ranging from dimensional regularization to real-time HTL computations. In addition, we discuss the mathematical properties of real-time propagators, the photon self-energy tensor and the low-momentum expansion, relevant for our work. A reader familiar with the topic may want to move directly to the following section.

\subsection{Conventions and notation}

We work in $D = 4 - 2\epsilon$ spacetime dimensions and $d=D-1$ spatial dimensions with the Minkowskian metric $g_{\mu\nu} = \mathrm{diag}(-1,+1,...,+1)$ and with the fermion Clifford algebra defined by $\big\{ \gamma^\mu,\gamma^\nu\big\} = -2g^{\mu\nu} $. All four-vectors are denoted by upper case letters and the magnitudes of spatial vectors with lower case letters,
\begin{equation}
\Pc \equiv (p^0,\mathbf{p}) \,, \quad p \equiv |\mathbf{p}| \,,
\end{equation}
where the individual spatial components are $p^i$, $i=1,...,d$. The $D$-dimensional integration measure is defined as
\begin{equation}
\int_{\Pc} \equiv \left (\frac{\exe^{\gamE}\overline{\Lambda}^2}{4\pi}\right )^{\frac{4-D}{2}} \int \frac{\mathrm{d}^D \Pc}{(2 \pi)^D} = \int_{-\infty}^\infty \frac{\mathrm{d}p^0}{2 \pi} \int_\mathbf{p} \,,
\end{equation}
where the spatial part of the integration measure can be written as \cite{Laine:2016hma}:
\begin{equation}
\begin{split}
\int_\mathbf{p} & \equiv \left (\frac{\exe^{\gamE}\overline{\Lambda}^2}{4\pi}\right )^{\frac{3-d}{2}} \int \frac{\mathrm{d}^d \mathbf{p}}{(2 \pi)^d}\\
& = \frac{4}{(4 \pi)^\frac{d+1}{2}\Gamma\left(\frac{d-1}{2}\right)} \left (\frac{\exe^{\gamE}\Lbar^2}{4\pi}\right )^{\frac{3-d}{2}}\int_0^\infty \mathrm{d}p\, p^{d-1} \int_{-1}^1 \mathrm{d}z (1-z^2)^\frac{d-3}{2} \,.
\end{split}
\label{eq:intmeasure}
\end{equation}
Here, the variable $z=\hat{\mathbf{k}}\cdot\hat{\mathbf{p}}$ parametrizes an angle with respect to some external vector $\mathbf{k}$ and $\Lbar$ is the $\overline{\text{MS}}$ renormalization scale. The factor $(\exe^{\gamE}/4\pi)^{(3-d)/2}$, with $\gamE$ the Euler--Mascheroni constant, is introduced as usual to simplify the final expressions.

\subsection{Real-time formalism}
\label{sec:realtimeform}

We calculate the self-energies using the $r/a$ (or Keldysh) basis representation of the real-time formalism (for a recent review see e.g.~\Ref\cite{Ghiglieri:2020dpq}), where propagators and self-energies are $2\times 2$ matrices
\begin{equation}
\mathbf{D} = \begin{pmatrix}
D^{rr} & D^R \\
D^A & 0 \\
\end{pmatrix} \,, \qquad
\mathbf{\Pi} = \begin{pmatrix}
0 & \Pi^A \\
\Pi^R & \Pi^{aa} \\
\end{pmatrix} \,,
\end{equation}
respectively. In terms of the $r/a$ indices, the retarded/advanced propagator is written as $D^{R/A} = D^{ra/ar}$, while for the self-energy, the corresponding result reads $\Pi^{R/A}=\Pi^{ar/ra}$.  In Feynman gauge ($\xi = 1$), the gauge-boson propagators are then given by 
\begin{equation}
D^{R/A}_{\mu\nu} (P) = g_{\mu\nu} \Delta^{R/A} (P) \,, \quad D^{rr}_{\mu\nu} (P) = g_{\mu\nu} \Delta^{rr}_B (P) \,,
\end{equation}
and the fermion propagators by
\begin{equation}
S^{R/A} (P) = -\slashed{P} \Delta^{R/A} (P) \,, \quad S^{rr} (P) = -\slashed{P} \Delta^{rr}_F (P) \,,
\end{equation}
where the retarded and advanced scalar parts are written as
\begin{equation}
\Delta^{R/A} (P) = \frac{-i}{P^2 \mp i \eta p^0} 
\end{equation}
with $\eta > 0$. 

The scalar $rr$-propagator is related to the retarded and advanced ones through the Kubo-Martin-Schwinger (KMS) relation,
\begin{equation}
\Delta^{rr}_{B} (P) = N_B (P) \Delta^d (P) \,, \quad \Delta^{rr}_{F} (P) = N_F^- (P) \Delta^d (P) \,,
\end{equation}
where the functions $N_B$ and $N_F$ are written in terms of the bosonic and fermionic distribution functions $n_{B/F} (p^0) = (\exe^{p^0/T}\mp 1)^{-1}$ as
\begin{equation}\label{eq:distfuncs}
N_B (P) = \frac{1}{2} + n_B(p^0) \,, \quad N_F^\pm (P) = \frac{1}{2} - n_F(p^0\pm\mu) \,,
\end{equation}
respectively. Here, $\Delta^d$ is the spectral function defined as the difference of the retarded and advanced propagators, and may be further expressed as $\delta$-functions by using the Sokhotski–Plemelj formula,
\begin{equation}\label{eq:scalarspectralfunc}
\Delta^d(P) = \Delta^R-\Delta^A = 2\pi \, \mathrm{sgn}(p^0) \delta(P^2) = \frac{\pi}{p} \big(\delta(p-p^0)-\delta(p+p^0)\big) \,.
\end{equation}

\begin{figure}[t!]
    \centering
    \includegraphics{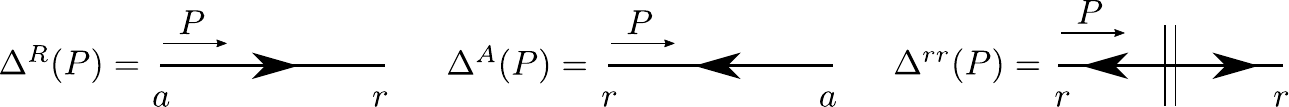}
    \caption{The $r/a$ propagators may be represented as arrows describing the flow of causality from $r$ fields to $a$ fields. The cut in the $rr$-propagator sources the flow.
    }
    \label{fig:rapropagators}
\end{figure}

In the computations to follow, we frequently rely on the parity properties of the above functions,
\begin{equation}\label{eq:funcparity}
\begin{alignedat}{3}
\Delta^A(\Pc) &= \Delta^R(-\Pc) \,, &\qquad \Delta^d(\Pc) &= -\Delta^d(-\Pc) \,, \\
  N_B^{ }(\Pc) &=-N_B^{ }(-\Pc) \,, &\qquad N_F^\pm(\Pc) &= -N_F^\mp(-\Pc) \,.
\end{alignedat}
\end{equation}
The different propagators in the $r/a$ basis may be interpreted as describing the flow of causality (see~\Ref\cite{Ghiglieri:2020dpq}). This interpretation leads to an intuitive graphical representation of the various $r/a$ assignments contributing to a certain diagram. In particular, the propagators are drawn as causal arrows from $a$ fields to $r$ fields as shown in~\fig\ref{fig:rapropagators}.

\begin{figure}[t!]
    \centering
    \includegraphics{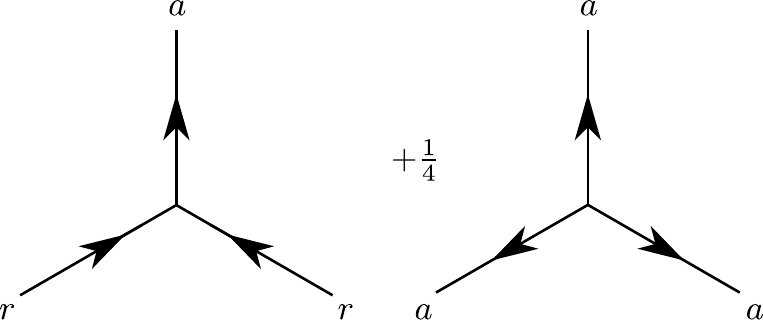}
    \caption{Two possible $r/a$ labelings for the three-point vertex. The vertex with three $a$ indices has an extra factor of $\frac{1}{4}$.}
    \label{fig:ra3vertices}
\end{figure}

In the $r/a$ basis, there are two distinct ways to assign $r/a$ labels to three-point vertices (relevant in QED), namely $rra$ and $aaa$. Additionally, the latter vertices are multiplied with an extra factor of $\frac{1}{4}$. Using the graphical causal arrow representation we introduced for the propagators, the two vertices are drawn in \fig\ref{fig:ra3vertices}.

\subsection{Tensor representation}\label{sec:tensorrep}
In QED, one defines the self-energy (or polarization tensor)  $\Pi^{\mu\nu}$ of the photon field through the Dyson-Schwinger equation as
\begin{equation}
\label{eq:SEdef}
i\Pi^{\mu\nu}(K) = (D^{-1})^{\mu\nu}(K) - (D^{-1}_0)^{\mu\nu}(K), 
\end{equation}
where $D^{\mu\nu}(K)$ is the full dressed photon propagator and $D^{\mu\nu}_{0}(K)$ is the bare propagator. The definition above implies that the self-energy $\Pi^{\mu\nu}$ is given as $+i$ times the appropriate Feynman diagram. The current conservation in QED requires that the photon self-energy is transverse
\begin{equation}
\label{eq:SEtransverse}
K_{\mu}\Pi^{\mu\nu}(K) = 0,    
\end{equation}
and gauge invariance requires that in a covariant gauge
\begin{equation}
\begin{split}
K_{\mu} K_{\nu}D^{\mu\nu}(K) = -i\xi,
\end{split}
\end{equation}
where the parameter $\xi$ fixes the gauge. In QED, both of these constraints hold in the vacuum as well as in medium \cite{Weldon:1996kb}. 

Introducing a thermal medium breaks the Lorentz symmetry of the vacuum by specifying a special frame of reference, the rest frame of the thermal bath. In that frame,
the remaining symmetry is associated with spatial rotations, and the four-velocity of the medium has the form $u^{\mu} = (1, \zerot)$. Consequently, the tensor basis for the photon self-energy extends to four different tensors $g_{\mu\nu}, K_\mu K_\nu, u_\mu u_\nu$ and $u_\mu K_\nu + u_\nu K_\mu$. Further, requiring the transversality property \eqref{eq:SEtransverse} reduces the number of available independent basis tensors to two, allowing us to write%
\footnote{Note that in QCD the transversality \nr{eq:SEtransverse} does not generally hold beyond the leading order HTLs~\cite{Weldon:1996kb}. Hence, the corresponding decomposition for the gluon includes all four basis tensors.}
\begin{equation}
\label{eq:SEsplitdef}
    \Pi^{\mu\nu}(K) = \prjoT^{\mu\nu}(K) \Pi_\text{T}(K) + \prjoL^{\mu\nu}(K) \Pi_\text{L}(K),
\end{equation}
where the scalar functions $\Pi_{\text{T}}$ and $\Pi_\text{L}$ are the transverse and longitudinal components of the self-energy tensor, respectively. The associated orthogonal projection operators read
\begin{equation}
\label{eq:projsDefs1}
\begin{split}
\prjoT^{\mu\nu}(K) & = \delta^{\mu}_{ i}\delta^{\nu}_{j}\biggl (g^{ij} - 
\frac{k^{i} k^{j}}{k^2} \biggr ),\\
\prjoL^{\mu\nu}(K) & = \prjof^{\mu \nu}(K)  - \prjoT^{\mu\nu}(K), 
\end{split}
\end{equation}
with
\begin{equation}
\label{eq:projsDefs2}
   \prjof^{\mu\nu}(K) = g^{\mu\nu} - \frac{K^{\mu}K^{\nu}}{K^2}.
\end{equation}
These above projectors are $D$-dimensionally transverse as required by \eq\eqref{eq:SEtransverse}, and additionally $\prjoT^{\mu\nu}(K)$ is $d$-dimensionally transverse with respect to $\kt$.

Using the properties of the two projectors, we can determine the coefficients seen in \eq\eqref{eq:SEsplitdef} via the trace and 00-component of the full self-energy tensor. The explicit $d$-dimensional identities are given by 
\begin{equation}
\label{eq:piTandpiL}
\begin{split}
\Pi_\text{T}(K) & = \frac{1}{d-1} \biggl (\Pi^{\mu}_{\mu}(K)  + \frac{K^2}{k^2} \Pi^{00}(K)\biggr ),\\
\Pi_\text{L}(K)& =  -\frac{K^2}{k^2} \Pi^{00}(K).
\end{split}
\end{equation}

\subsection{HTL limit}
\label{sec:HTLlimit}

In calculations involving a thermal medium, it is often convenient to extract the vacuum contribution from the self-energy tensor as 
\begin{equation}
\Pi^{\mu\nu} = \left(\Pi^{\mu\nu} - \lim_{T,\mu\to 0}\Pi^{\mu\nu} \right) + \lim_{T,\mu\to 0}\Pi^{\mu\nu} \equiv \Pi^{\mu\nu}_{\rm M} +  \Pi^{\mu\nu}_{\rm V} \,,
\end{equation}
where the vacuum-subtracted $\Pi^{\mu\nu}_{\rm M}$ is the matter contribution and $\Pi^{\mu\nu}_{\rm V}$ is the ultraviolet (UV) -divergent vacuum contribution independent of the medium. However, when considering the self-energy at small external momentum $K$, it is beneficial to write the small-$K$ expansion without explicitly separating the vacuum and matter parts,
\begin{equation}\label{eq:smallKexpansion}
\Pi^{\mu\nu}(K) = \mE^2 \sum_{n=0}^\infty \left(\frac{\che^2K^2}{\mE^2} \right)^n c^{\mu\nu}_{n}(\mu/T,k^0/k) \,,
\end{equation}
where $\che$ is the electric charge, $\mE^2$ is an $O(\che^2)$  effective thermal mass scale (to be defined below), and the $c^{\mu\nu}_n$ are dimensionless functions.%
\footnote{Note that in general they can also depend on ``mixed'' ratios of thermal and momentum scales. However, for the power corrections the angular and radial integrals are expected to factorise in such a way that this does not happen. In the one-loop case this is evident from the calculations performed in this paper.}
Since $\Pi^{\mu\nu}_{\rm V} \sim K^2$ by Lorentz symmetry, only $\Pi^{\mu\nu}_{\rm M}$ contributes to the $n=0$ term. It also follows that the $n=1$ term is a sum of $\Pi^{\mu\nu}_{\rm V}$ and the $O(K^2)$ piece of $\Pi^{\mu\nu}_{\rm M}$, each of which separately contains terms proportional to $\log(K^2)$, but in the sum they cancel out, leaving only ratios of mass scales $\log(T/\Lbar)$ and $\log(\mu/\Lbar)$. (Here and throughout, $\log$ denotes the complex logarithm). Further, it is worth noting that the structure of \eq\eqref{eq:smallKexpansion} implies that the $n=1$ term (unless $c_1^{\mu\nu}$ has a very particular form) includes nontrivial structures originating from $\Pi^{\mu\nu}_{\rm M}$ that are nevertheless independent of $T$ and $\mu$.

In this work, we compute the one-loop and two-loop parts of the $n=0$ term of \eq\eqref{eq:smallKexpansion}, of which the former is the usual LO HTL self-energy, as well as the one-loop part of the $n=1$ term, dubbed a ``power correction''. The contributions are denoted by $\Pi_\mathrm{LO}$, $\Pi_\mathrm{NLO}$ and $\Pi_\mathrm{Pow}$, respectively.

\section{Detailed evaluation of the photon self-energies}
\label{sec:detailedcomputation}
\noindent
In this section, we present the detailed computation of $\Pi_\mathrm{LO}$, $\Pi_\mathrm{Pow}$, and $\Pi_\mathrm{NLO}$ with arbitrary soft external momenta in a hot (nonzero-temperature) and dense (nonzero-chemical-potentials) QED plasma, generalizing the results of  \cite{Manuel:2016wqs,Carignano:2017ovz,Carignano:2019ofj} to nonzero density.  The results of the computation are collected in their entirety in the following section. However, the results follow the notation defined in this section, so we encourage the reader to review also the present section at least cursorily.

The calculations are performed within the standard $r/a$-basis representation of the real-time formalism. We work in the massless-fermion limit and, as discussed above, use dimensional regularization to regularize all the intermediate singularities in the $\overline{\text{MS}}$ renormalization scheme.

\subsection{One-loop photon self-energy}

\begin{figure}[t!]
    \centering
    \includegraphics{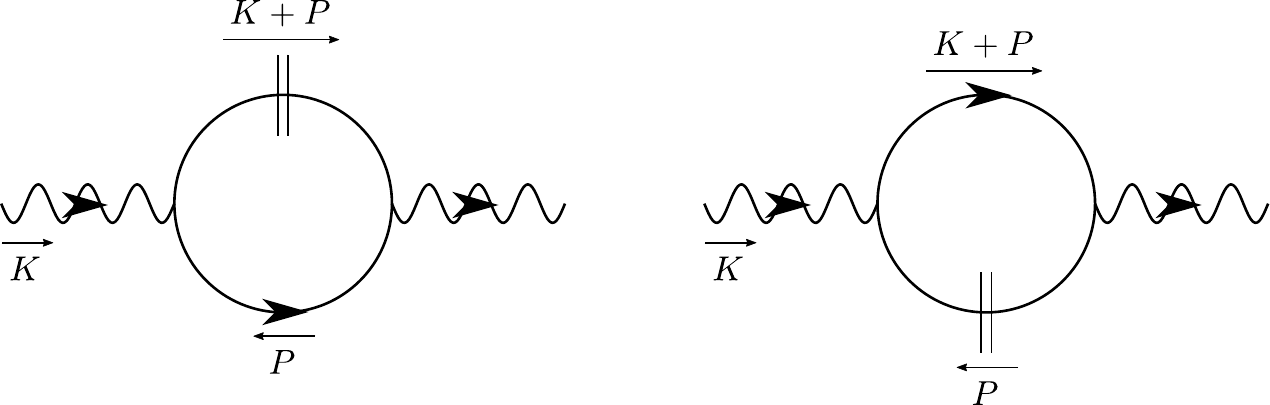}
    \caption{The two $r/a$ assignments for the retarded one-loop photon self-energy diagram. By convention, the direction of fermion flow is aligned with the momenta of fermions.}
    \label{fig:photon_1loop}
\end{figure}

There are two ways to draw the $r/a$ arrows to a retarded one-loop photon self-energy as shown in \fig\ref{fig:photon_1loop}. An application of Feynman rules to the assignments yields
\begin{align}
-i (\Pi^R_\mathrm{LO})_{\mu \nu} (\Kc) = -\int_\Pc F_{\mu\nu}(\Kc,\Pc) \bigg\{\Delta^A(\Pc) \Delta^{rr}_F(\Kc+\Pc) + \Delta^{rr}_F(\Pc) \Delta^R(\Kc+\Pc)\bigg \} \,, \label{eq:quarkbubble1}
\end{align}
where the overall minus sign stems from the fermion loop. The numerator algebra is contained in the tensor $F_{\mu\nu}$ and simplifies to 
\begin{align}
F_{\mu\nu}(\Kc,\Pc) &= \mathrm{Tr}\left[\left(iV_{\mu}\right) \slashed{\Pc} \left(iV_{\nu}\right) (\slashed{\Kc}+\slashed{\Pc}) \right] \nonumber \\
&= -4\che^2 \biggl ( 2\Pc_\mu \Pc_\nu + \Kc_\mu \Pc_\nu + \Pc_\mu \Kc_\nu - \left(\Pc^2+\Kc \cdot \Pc\right)g_{\mu\nu} \biggr ) \,,
\label{eq:quarknumerator}
\end{align}
where $iV_\mu = i \che \gamma_\mu$ stands for the free electron--photon vertex function. 

Next, we should bring \eq\eqref{eq:quarkbubble1} to a form where the $\delta$-functions and distribution functions in the $rr$-propagators depend only on the loop momentum $\Pc$, so that the distribution functions become independent of the angle between $\mathbf{k}$ and $\mathbf{p}$, making the angular integral easier to handle. To this end, we shift the loop momentum in the first term by $\Pc \mapsto -\Kc-\Pc$. The numerator is invariant under this change of variables due to the cyclicity of the trace and the symmetry under $\mu \leftrightarrow \nu$. Hence, we obtain
\begin{equation}
-i (\Pi^R_\mathrm{LO})_{\mu \nu} (\Kc) = -\int_\Pc F_{\mu\nu}\bigg \{N_F^-(\Pc)+N_F^+(\Pc)\bigg \} \Delta^R(\Kc+\Pc) \Delta^d(\Pc) \,, 
\label{eq:quarkbubble2}
\end{equation}
where we utilized the parity properties of the propagators and distribution functions we introduced in \eq\eqref{eq:funcparity}. From the expression in \eq\eqref{eq:quarkbubble2}, we can then compute the trace $(\Pi^R_\mathrm{LO})^\mu_\mu$ and 00-component $(\Pi^R_\mathrm{LO})_{00}$ of the photon self-energy.

\subsubsection{Trace}

Contracting the metric tensor with the numerator in \eq\eqref{eq:quarknumerator} gives
\begin{align}
F_\mu^\mu &= 4\che^2 (D-2) \Kc \cdot \Pc \,,
\end{align}
where the spectral function $\Delta^d$ sets $\Pc$ on-shell, $\Pc^2=0$. Substituting the above numerator into \eq\eqref{eq:quarkbubble2} and writing the $\Delta^R$ propagator out explicitly yields the result
\begin{equation}
\begin{split}
(\Pi^R_\mathrm{LO})_\mu^\mu (\Kc) = -4\che^2(D-2)\int_\Pc \Delta^d(\Pc) \bigg\{N_F^-(\Pc)+N_F^+(\Pc)\bigg\} \frac{\Kc \cdot \Pc}{2\Kc \cdot \Pc + \Kc^2} \,. 
\label{eq:g1looptrace}
\end{split}
\end{equation}
To avoid clutter, the $i\eta$ from the retarded propagator has been absorbed into $k^0$ since we can write $(\Kc+\Pc)^2 -i\eta(k^0+p^0) = (\Kc+\Pc)^2|_{k^0\to k^0+i\eta}$ for small $\eta$. Hence, later on we should remember that the 0-component of the external momentum has a small imaginary part and replace $k^0$ with $k^0+i\eta$ in our expressions.

In the HTL limit, we are interested in the behavior of the photon self-energy when the external momentum $K$ is soft. A convenient shortcut for extracting the limit is to expand the integrand for $\Kc \ll \mathbf{p}$. Strictly speaking, we should first integrate over $p^0$ properly, but since $\Delta^d$ is proportional to $\delta(p^0-p) - \delta(p^0+p)$, it has already been done implicitly by the $\delta$-functions. Before expanding, we should therefore set $\Pc^2=0$ to obtain the correct expansion. Hence, we can directly expand \eq\eqref{eq:g1looptrace} for small $\Kc$ and pick out the leading terms, yielding
\begin{equation}
\begin{split}
(\Pi^R_\mathrm{LO})_\mu^\mu (\Kc) = -2\che^2(D-2)\int_\Pc \Delta^d(\Pc) \bigg\{ N_F^-(\Pc)+N_F^+(\Pc) \bigg\} \,. 
\label{eq:g1looptrace2}
\end{split}
\end{equation}
In general, expressions become more symmetric in the HTL limit, which is the reason we have taken the limit before integrating explicitly. In our case, the expression in the curly brackets in \eq\eqref{eq:g1looptrace2} is odd in $\Pc$, so we can carry out the $p^0$-integral by employing the formula
\begin{equation}
\int_\Pc \Delta^d(\Pc) f(p^0,\pt) = \int_\mathbf{p} \frac{f(p,\pt)}{p} \,, \label{eq:p0intformula}
\end{equation}
where $f$ is an odd function of $\Pc$ ($\Delta^d$ is odd, making the integrand even). This is a direct consequence of the definition of $\Delta^d$ in \eq\eqref{eq:scalarspectralfunc}. 

Integrating over $p^0$ gives next
\begin{align}
(\Pi^R_\mathrm{LO})_\mu^\mu (\Kc) &= -2\che^2(D-2) \int_\mathbf{p} \frac{1}{p} \bigg\{N_F^-(p)+N_F^+(p)\bigg\} \nn
&= -2\che^2 (d-1) \mathcal{N} \mathcal{R}_2 \mathcal{A}_0 \equiv \mE^2 \,, 
\label{eq:g1looptrace3}
\end{align}
where $N_F^\pm(p)$ is understood as $N_F^\pm(\Pc)|_{p^0\to p}$. Characteristic to the HTL expansion, the radial and angular spatial integrals factorize making their separate calculation possible. Here, we have introduced the notation of \app\ref{app:integrals}, where the radial integrals are denoted by $\mathcal{R}_i$ and angular ones by $\mathcal{A}_i$. The normalization of the integration measure given by \eq\eqref{eq:intmeasure} has been absorbed into the factor
\begin{equation}
\mathcal{N} \equiv \frac{4}{(4 \pi)^\frac{d+1}{2}\Gamma\left(\frac{d-1}{2}\right)} \left (\frac{\exe^{\gamE}\Lbar^2}{4\pi}\right )^{\frac{3-d}{2}} \,.\label{eq:intnormfactor}
\end{equation}
We also defined $\mE$ as the $d$-dimensional in-medium effective mass scale for the photon. 

Next, we apply the results for the spatial integrals found in \app\ref{app:integrals}. The integrals are regulated by dimensional regularization in $d$ spatial dimensions but neither in $\mathcal{R}_2$ nor $\mathcal{A}_0$ divergences are present. To obtain the finite part, we may then set $d=3$ everywhere leading to the result
\begin{equation}
\label{eq:Dmassd3}
(\Pi^R_\mathrm{LO})_\mu^\mu (\Kc) = \che^2 \bigg (\frac{T^2}{3} + \frac{\mu^2}{\pi^2} \bigg ) = \mE^2\bigr|_{d=3} \,,
\end{equation}
which gives the well-known $d=3$-dimensional value for $\mE$.

\subsubsection{00-component}
Next, we repeat the above steps for the 00-component. Picking out the 00-component from the numerator in \eq\eqref{eq:quarknumerator} gives
\begin{align}
F_{00} = -4\che^2 \biggl ( 2p_0^2 + 2 k^0 p^0 + \Kc \cdot \Pc \biggr ) \,,
\end{align}
and inserting this expression into \eq\eqref{eq:quarkbubble2} leads to
\begin{equation}
\begin{split}
(\Pi^R_\mathrm{LO})_{00} (\Kc) = 4\che^2\int_\Pc \Delta^d(\Pc) \bigg\{N_F^-(\Pc)+N_F^+(\Pc)\bigg\} \frac{2p_0^2 + 2 k^0 p^0 + \Kc \cdot \Pc}{2\Kc \cdot \Pc + \Kc^2} \,. 
\label{eq:g1loop00}
\end{split}
\end{equation}
The HTL limit is obtained by expanding in small $\Kc$, resulting in
\begin{equation}
\begin{split}
(\Pi^R_\mathrm{LO})_{00} (\Kc) = 2\che^2\int_\Pc \Delta^d(\Pc) \bigg\{N_F^-(\Pc)+N_F^+(\Pc) \bigg\} \left(1 + \frac{2k^0 p^0}{\Kc\cdot\Pc} - \frac{\Kc^2 p_0^2}{(\Kc\cdot\Pc)^2} + \frac{2p_0^2}{\Kc\cdot\Pc} \right) \,.
\end{split}
\end{equation}
The last term in the round brackets behaves parametrically as $O(1/\Kc)$ and leads the small-$\Kc$ expansion. However, the term vanishes due to symmetry since the corresponding integrand is an odd function of the $D$-dimensional vector $\Pc$. The remaining terms become leading, and since their integrands are even, we can use \eq\eqref{eq:p0intformula} to integrate over $p^0$,
\begin{align}
(\Pi^R_\mathrm{LO})_{00} (\Kc) &= 2\che^2\int_\mathbf{p} \frac{1}{p} \bigg\{ 
N_F^-(p)+N_F^+(p)\bigg\} \left(1 + \frac{2k^0}{v\cdot\Kc} - \frac{\Kc^2}{(v\cdot\Kc)^2} \right) \nn
&= 2\che^2 \mathcal{N} \mathcal{R}_2 \left(\mathcal{A}_0 + 2k^0 \mathcal{A}_1 - K^2 \mathcal{A}_2 \right) \,, \label{eq:g1loop001}
\end{align}
where $v \equiv (1,\hat{\mathbf{p}})$ so that $v\cdot\Kc = -k^0+kz$. The radial and angular integrals in \eq\eqref{eq:g1loop001} are finite,
allowing us to set $d=3$, which leads to
\begin{equation}
\label{eq:g1loop001final}
(\Pi^R_\mathrm{LO})_{00} (\Kc) = -\mE^2\bigr|_{d=3}\bigg(1 - k^0 L(\Kc) \bigg ) \,,
\end{equation}
with
\begin{equation}
L(\Kc) \equiv \frac{1}{2k} \log \frac{k^0+k}{k^0-k} \,. \label{eq:HTLlogarithm}
\end{equation}

\subsection{Power corrections to one-loop photon self-energy}
\label{sec:powcor}

Next, we will consider the leading power correction to the one-loop photon self-energy at nonzero temperature and density. As in the leading-order HTL case, we start from the retarded photon self-energy tensor in \eq\nr{eq:quarkbubble2} and consider the kinematical approximation where the external momentum is soft $K \ll P$. As discussed in \Sec\ref{sec:HTLlimit} (see \eq\nr{eq:smallKexpansion}), the power corrections are then obtained by expanding the argument of the one-loop photon self-energy in powers $(K^2)^n$, where $n=0$  gives the leading HTL result and $n=1$ corresponds to the power correction.

Concentrating first the trace part of the self-energy tensor, we obtain for the first subleading $(\sim K^2)$ power correction term
\begin{equation}
\begin{split}
(\Pi^R_{\rm Pow})_\mu^\mu (\Kc) = -\frac{\che^2}{2}(D-2)\int_\Pc \Delta^d(\Pc) \bigg\{ N_F^-(\Pc)+N_F^+(\Pc)\bigg\} \frac{K^4}{(\Kc \cdot \Pc)^2} \,, 
\label{eq:g1looptracepowcor}
\end{split}
\end{equation}
where the integral over $p^0$ yields
\begin{align}
(\Pi^R_{\rm Pow})_\mu^\mu (\Kc) & = -\frac{\che^2}{2}(D-2)\int_\mathbf{p} \frac{1}{p^3}\bigg\{ N_F^-(p)+N_F^+(p)\bigg\} \frac{K^4}{(v \cdot \Kc)^2} \nn
& = -\frac{\che^2}{2}(d-1)K^4 \mathcal{N} \mathcal{R}_3 \mathcal{A}_2 \,.
\label{eq:g1looptracepowcor2}
\end{align}

Next, we concentrate on the 00-component of the self-energy tensor, for which the leading power correction term is given by
\begin{equation}
\begin{split}
(\Pi^R_{\rm Pow})_{00}(\Kc) = -\frac{\che^2}{2}\int_\Pc \Delta^d(\Pc) \bigg\{N_F^-(\Pc)+N_F^+(\Pc) \bigg\}\! \left (\frac{K^6p_0^2}{(\Kc \cdot \Pc)^4} - \frac{K^4(2k^0 p^0 + \Kc \cdot \Pc)}{(\Kc \cdot \Pc)^3}\right ) \,,
\label{eq:g1loop00powcor}
\end{split}
\end{equation}
where the integral over $p^0$ yields
\begin{align}
(\Pi^R_{\rm Pow})_{00}(\Kc) & = -\frac{\che^2}{2}\int_\mathbf{p} \frac{1}{p^3} \bigg\{N_F^-(p)+N_F^+(p) \bigg\} \left (\frac{K^6}{(v \cdot \Kc)^4} - \frac{2k^0 K^4}{(v \cdot \Kc)^3} - \frac{K^4}{(v \cdot \Kc)^2}\right ) \nn
& = -\frac{\che^2}{2} K^4 \mathcal{N} \mathcal{R}_3 \left( K^2\mathcal{A}_4 - 2k^0 \mathcal{A}_3 - \mathcal{A}_2 \right) \,.
\label{eq:g1loop00powcor2}
\end{align}

Here, the radial and angular integrals factorize and are computed up to $O(\varepsilon)$ in \app\ref{app:integrals}. Note that the radial integral $\mathcal{R}_3$ is proportional to
\begin{equation}
\label{eq:uvdivpow}
\begin{split}
\int_0^\infty  \frac{\ud p \,p^{d-1}}{p^3} \bigg \{1 - n_F(p - \mu) - n_F(p+\mu)\biggr \} \,,
\end{split}
\end{equation}
which is divergent when $d \to 3$. Since the internal momentum is large compared to the external soft momentum $K$, the associated divergence appearing as a $1/\varepsilon$ term is a UV divergence of full QED, rather than of the HTL theory (the latter of which would arise to compensate for an IR divergence of the full theory).%
\footnote{In fact, the UV divergence originates from the vacuum self-energy $\Pi_\mathrm{V}$ (see the discussion in \Sec\ref{sec:HTLlimit}).}
Consequently, the divergence seen here will be cancelled upon the UV renormalization of the full theory. The structure of the divergences is discussed more extensively in the context of the pressure in \Ref\cite{Gorda:2021kme}.

Using the results given in \app\ref{app:integrals}, we may finally write
\begin{equation}
\label{eq:powcortr}
\begin{split}
(\Pi^R_{\rm Pow})_\mu^\mu (K) & = -\frac{\che^2 K^2}{2\pi^2} \bigg \{-\frac{1}{2\varepsilon} +  \left (\log \frac{2\exe^{-\gamE}T}{\Lbar} - 1 \right ) + \frac{1}{2}\\
& + \left (1 - \frac{k^0}{2k} \log\frac{k^0+k+i\eta}{k^0-k+i\eta} \right ) - \mathrm{Li}^{(1)}_0(-\exe^{\frac{\mu}{T}}) - \mathrm{Li}^{(1)}_0(-\exe^{-\frac{\mu}{T}}) \biggr \} \,,
\end{split}    
\end{equation}
and
\begin{equation}
\label{eq:powcor00}
\begin{split}
(\Pi^R_{\rm Pow})_{00}(\Kc)  & = \frac{\che^2 k^2}{6\pi^2} \bigg \{-\frac{1}{2\varepsilon} + \left (\log \frac{2\exe^{-\gamE}T}{\Lbar} - 1 \right )\\
& + \frac{1}{2}\left (3 - \frac{k_0^2}{k^2} \right ) \left (1 - \frac{k^0}{2k} \log\frac{k^0+k+i\eta}{k^0-k+i\eta} \right ) - \mathrm{Li}^{(1)}_0(-\exe^{\frac{\mu}{T}}) - \mathrm{Li}^{(1)}_0(-\exe^{-\frac{\mu}{T}}) \biggr \} \,,
\end{split}    
\end{equation}
where we have introduced the notation $\mathrm{Li}^{(1)}_0(z) = \lim_{s \to 0}\frac{\partial \mathrm{Li}_s(z)}{\partial s}$ with $\mathrm{Li}_s$ standing for the standard polylogarithm function.  As mentioned above, the UV divergence in these expressions is  eliminated with UV renormalization, specifically by renormalizing the wave function. In the $\overline{\mathrm{MS}}$ scheme, the wave-function renormalization constant $Z_3$ for the photon field reads $Z_3 = 1 - \frac{\che^2}{4\pi^2} \frac{1}{3\varepsilon}$. This leads to the renormalized expressions which will be presented in Section~\ref{sec:results}.

\subsection{Two-loop photon self-energy}

We are now finally able to move on to a detailed presentation of the calculation of the retarded two-loop photon self-energy in the HTL limit, which generalizes the calculation of Carignano et al.~\cite{Carignano:2019ofj} to finite density. We have explicitly checked that the $\mu\to 0$ limit of $\Pi^R_\mathrm{NLO}$ agrees with the result of \Ref\cite{Carignano:2019ofj}. 

There are three diagrams that contribute to the self-energy at the two-loop order, which we call the cat's eye graph, watermelon~1 (denoted by M1) and watermelon~2 (M2) 
\begin{equation}
(\Pi^R_\mathrm{NLO})_{\mu \nu} = (\Pi^R_\mathrm{cat})_{\mu \nu} + (\Pi^R_\mathrm{M1})_{\mu \nu} + (\Pi^R_\mathrm{M2})_{\mu \nu} \,, \label{eq:2loopcontribs}
\end{equation}
and which are represented by the  graphs%
\footnote{As above, the direction of the fermionic flow is aligned with the momentum assignments.}
\begin{equation}\label{eq:2loopgraphs}
\raisebox{-0.5\height}{\includegraphics[scale=1.10]{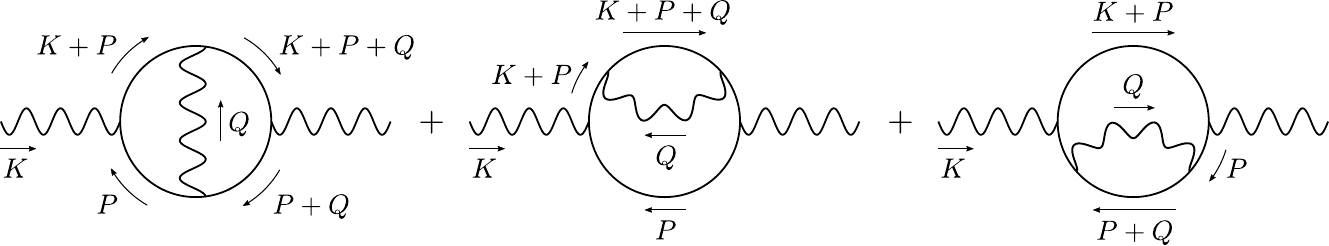}} \,,
\end{equation}
respectively.

Before moving to the actual computation, let us first present a useful relation between the distribution functions $N_B^{ }$ and $N_F^\pm$. At two-loop level, the integrals contain products of two distribution functions which obey the relation
\begin{equation}\label{eq:2distrel}
    N_F^\pm(\Pc_1)N_F^\mp(\Pc_2)
  + N_F^\mp(\Pc_2)N_B^{ }(\Pc_3)
  + N_B^{ }(\Pc_3)N_F^\pm(\Pc_1)+1=0 \,,
\end{equation}
where the momenta satisfy $\Pc_1+\Pc_2+\Pc_3=0$. Since the last constant term on the LHS in the above relation is a pure vacuum term not contributing to the HTL limit, it will be persistently ignored in the following.

\subsubsection{Cat's-eye diagram}

\begin{figure}[t!]
    \centering
    \includegraphics[width=\textwidth]{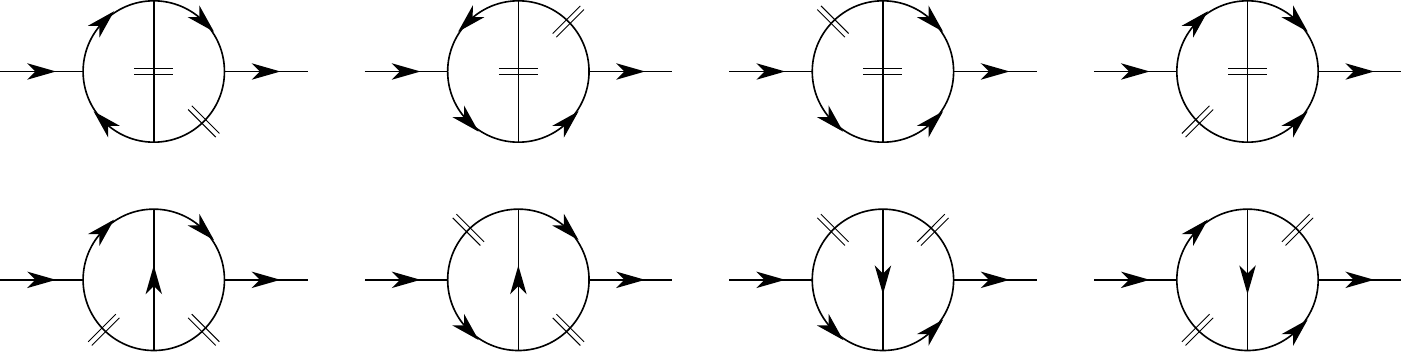}
    \caption{$r/a$ assignments contributing to the thermal parts of the cat's eye diagram. Here, the distinction between photon and fermion lines is not necessary since assigning the $r/a$ labels depends only on the topology of a given diagram.}
    \label{fig:ra_cat}
\end{figure}

Applying the Feynman rules to the cat's-eye diagram without specifying the $r/a$ assignments yields 
\begin{equation}
- \int_\Pc \int_\Qc \mathrm{Tr} \left[ \big(i V_\mu\big) S^p(\Pc) \big(i V_{\rho}\big) S^q(\Pc\Qc) \big(i V_\nu\big) S^r(\Kc\Pc\Qc) \big(i V_{\sigma}\big) S^s(\Kc\Pc) \right]  D^{\rho \sigma}_t(\Qc) \,,
\label{eq:cattemplate}
\end{equation}
where the indices $p,q,r,s$ and $t$ will be replaced with the appropriate $r/a$ labels. To save space, we have compactified the notation by suppressing plus signs in the arguments of propagators: $\Kc \Pc \Qc = \Kc + \Pc + \Qc$, etc. The overall minus sign originates from the fermion loop present in the diagram.
Applying the $r/a$-basis Feynman rules to the above expression yields now the assignments (see \fig\ref{fig:ra_cat}):
\begin{equation}
\begin{split}
-i (\Pi^R_\mathrm{cat})_{\mu \nu} (\Kc) &= - \int_\Pc \int_\Qc F^\mathrm{cat}_{\mu\nu} (\Kc,\Pc,\Qc) \\
\times \biggl\{ &\Delta^R(\Pc) \Delta_B^{rr}(\Qc) \Delta_F^{rr}(\Pc\Qc) \Delta^R(\Kc\Pc) \Delta^R(\Kc\Pc\Qc) \\
+&\Delta^A(\Pc) \Delta_B^{rr}(\Qc) \Delta^A(\Pc\Qc) \Delta^A(\Kc\Pc) \Delta_F^{rr}(\Kc\Pc\Qc) \\
+&\Delta^A(\Pc) \Delta_B^{rr}(\Qc) \Delta^A(\Pc\Qc) \Delta_F^{rr}(\Kc\Pc) \Delta^R(\Kc\Pc\Qc) \\
+&\Delta_F^{rr}(\Pc) \Delta_B^{rr}(\Qc) \Delta^A(\Pc\Qc) \Delta^R(\Kc\Pc) \Delta^R(\Kc\Pc\Qc) \\
+&\Delta_F^{rr}(\Pc) \Delta^R(\Qc) \Delta_F^{rr}(\Pc\Qc) \Delta^R(\Kc\Pc) \Delta^R(\Kc\Pc\Qc) \\
+&\Delta^A(\Pc) \Delta^R(\Qc) \Delta_F^{rr}(\Pc\Qc) \Delta_F^{rr}(\Kc\Pc) \Delta^R(\Kc\Pc\Qc) \\
+&\Delta^A(\Pc) \Delta^A(\Qc) \Delta^A(\Pc\Qc) \Delta_F^{rr}(\Kc\Pc) \Delta_F^{rr}(\Kc\Pc\Qc) \\
+&\Delta_F^{rr}(\Pc) \Delta^A(\Qc) \Delta^A(\Pc\Qc) \Delta^R(\Kc\Pc) \Delta_F^{rr}(\Kc\Pc\Qc) \biggr\} \,.
\end{split}
\label{eq:cat1}
\end{equation}
We have only included assignments with thermal contributions and dropped all purely vacuum terms, i.e.~those not containing any $rr$-propagators. Additionally, assignments with closed loops of retarded or advanced propagators have been discarded since they integrate to zero. 

The numerator algebra above simplifies to
\begin{align}
F^\mathrm{cat}_{\mu\nu} &= \mathrm{Tr} \left[ \big(i V_\mu\big) \slashed{\Pc} \big(i V_{\rho}\big) (\slashed{\Pc}+\slashed{\Qc}) \big(i V_\nu\big) (\slashed{\Kc}+\slashed{\Pc}+\slashed{\Qc}) \big(i V_{\sigma}\big) (\slashed{\Kc}+\slashed{\Pc}) \right]  g^{\rho \sigma} \nonumber \\
&= \che^4 \mathrm{Tr} \left[ \gamma_\mu \slashed{\Pc} \gamma^\rho (\slashed{\Pc}+\slashed{\Qc}) \gamma_\nu (\slashed{\Kc}+\slashed{\Pc}+\slashed{\Qc}) \gamma_\rho (\slashed{\Kc}+\slashed{\Pc}) \right] \,,
\end{align}
where the Dirac trace can be computed in $D$ dimensions by employing well-known $\gamma$-matrix identities. However, the resulting expression is rather lengthy and we choose not to write it down explicitly. As the trace and 00-component end up being the only components we need, we may utilize that the diagonal elements $(\mu=\nu)$ of $F^\mathrm{cat}_{\mu\nu}$ are invariant under the following changes of integration variables
\begin{align}
&(1) \quad \Pc \mapsto -\Kc-\Pc-\Qc \,, \\
&(2) \quad \Pc \mapsto -\Kc-\Pc \,, \quad \Qc \mapsto -\Qc \,.
\end{align}

Next, our goal is to make the integration more manageable by bringing \eq\eqref{eq:cat1} into a form where the $\Delta^d$ functions and distribution functions depend only on single momenta. This is significantly more complicated than in the one-loop case, but can be achieved using the relation between the distribution functions in \eq\eqref{eq:2distrel}. That relation, together with the variable changes (1) and (2) allows us to simplify \eq\eqref{eq:cat1} to
\begin{equation}
\begin{split}
-i (\Pi^R_\mathrm{cat})_{\mu \nu} (\Kc) &= - \int_\Pc \int_\Qc F^\mathrm{cat}_{\mu\nu} (\Kc,\Pc,\Qc) \Delta^d(\Pc) \Delta^R(\Kc+\Pc) \\
  &\times \Big\{2 N_B^{ }(\Qc) \left[N_F^-(\Pc)+N_F^+(\Pc)\right] \Delta^R(\Kc+\Pc+\Qc) \\
&\qquad \qquad\times \big[\Delta^d(\Pc+\Qc)\Delta^p(\Qc)+\Delta^d(\Qc)\Delta^p(\Pc+\Qc)\big] \\
&\qquad+ \left[N_F^-(\Pc)N_F^-(\Kc+\Pc+\Qc)+N_F^+(\Pc)N_F^+(\Kc+\Pc+\Qc)\right] \\
&\qquad \qquad\times \Delta^A(\Qc) \Delta^A(\Pc+\Qc) \Delta^d(\Kc+\Pc+\Qc) \Big\} \,,
\end{split}\label{eq:cat2}
\end{equation}
where we utilized the parity properties of the functions listed in \Sec\ref{sec:realtimeform} and introduced the notation $\Delta^p \equiv(\Delta^R+\Delta^A)/2$. By introducing another two changes of variables,
\begin{align}
&(3) \quad \Qc \mapsto -\Kc-\Pc-\Qc \,, \\
&(4) \quad \Qc \mapsto \Qc-\Pc \,,
\end{align}
and making use of \eq\eqref{eq:2distrel} again, we can bring \eq\eqref{eq:cat2} into the form
\begin{equation}
\begin{split}
-&i (\Pi^R_\mathrm{cat})_{\mu \nu} (\Kc) = - \int_\Pc \int_\Qc \Delta^d(\Pc) \Delta^d(\Qc) \\
\times& \Big\{2\Delta^R(\Kc+\Pc) \Delta^p(\Pc+\Qc) \Delta^R(\Kc+\Pc+\Qc) N_B(\Qc) \left[N_F^-(\Pc)+N_F^+(\Pc)\right] F^{\mathrm{cat}}_{\mu\nu} \\
&+ \Delta^R(\Kc+\Pc) \Delta^R(\Kc+\Qc) \Delta^R(\Kc+\Pc+\Qc)  \left[N_F^-(\Pc)N_F^+(\Qc)+N_F^+(\Pc)N_F^-(\Qc)\right] F^{\mathrm{cat}(3)}_{\mu\nu} \\
&+ \Delta^R(\Kc+\Pc) \Delta^p(\Pc-\Qc) \Delta^R(\Kc+\Qc)  \left[N_F^-(\Pc)N_F^-(\Qc)+N_F^+(\Pc)N_F^+(\Qc)\right] F^{\mathrm{cat}(4)}_{\mu\nu} \Big\} \,,
\end{split}\label{eq:cat3}
\end{equation}
where the variable changes (3) and (4) have not left the numerator $F^\mathrm{cat}_{\mu\nu}$ invariant, and we have denoted the results of those changes by $F^{\mathrm{cat}(i)}_{\mu\nu}$, with $(i)$ referring to the particular change of variables.

\subsubsection{Watermelon diagram 1}

\begin{figure}[t!]
    \centering
    \includegraphics[width=\textwidth]{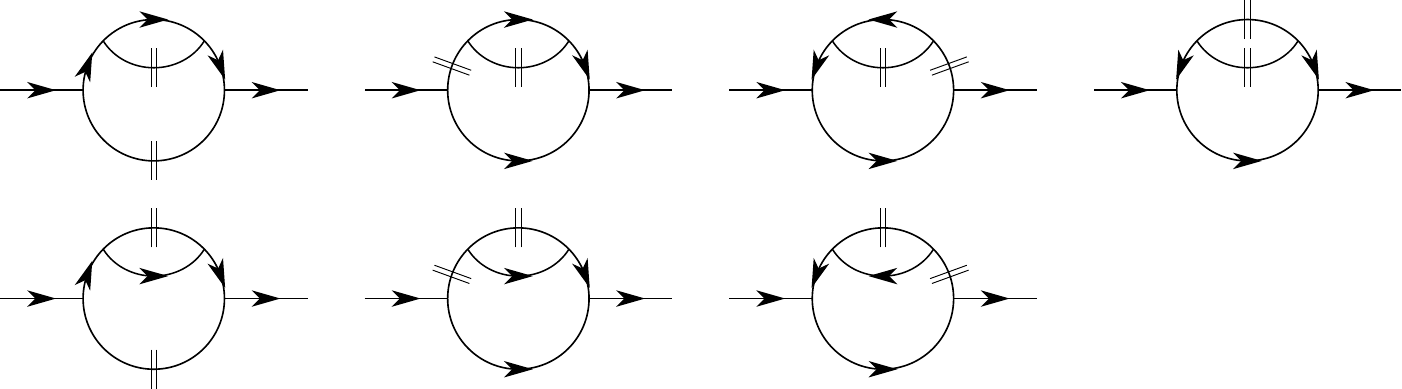}
    \caption{$r/a$ assignments contributing to the thermal parts of the watermelon diagram.}
    \label{fig:ra_melon}
\end{figure}

Next, we concentrate on the watermelon diagram 1 shown in \eq\nr{eq:2loopgraphs}. When applied to it,  the Feynman rules yield
\begin{equation}
- \int_\Pc \int_\Qc \mathrm{Tr} \left[ \big(i V_\mu\big) S^p(\Pc) \big(i V_\nu\big) S^q(\Kc\Pc) \big(i V_{\rho}\big) S^r(\Kc\Pc\Qc) \big(i V_{\sigma}\big) S^s(\Kc\Pc) \right]  D^{\rho \sigma}_t(\Qc) \,,
\end{equation}
to which we assign the appropriate $r/a$ labels as shown in \fig\ref{fig:ra_melon}. Again, contributions independent of distribution functions (i.e.~the vacuum parts) and contributions containing closed causality loops are neglected. This leads to the  expression
\begin{align}
-i (\Pi^R_\mathrm{M1})_{\mu \nu} (\Kc) &= - \int_\Pc \int_\Qc F^\mathrm{M1}_{\mu\nu} (\Kc,\Pc,\Qc) \nn
\times \biggl \{ &\Delta_F^{rr}(\Pc) \Delta_B^{rr}(\Qc) \Delta^R(\Kc\Pc) \Delta^R(\Kc\Pc) \Delta^R(\Kc\Pc\Qc) \nn
&+\Delta^A(\Pc) \Delta_B^{rr}(\Qc) \Delta_F^{rr}(\Kc\Pc) \Delta^R(\Kc\Pc) \Delta^R(\Kc\Pc\Qc) \nn
&+\Delta^A(\Pc) \Delta_B^{rr}(\Qc) \Delta_F^{rr}(\Kc\Pc) \Delta^A(\Kc\Pc) \Delta^A(\Kc\Pc\Qc) \nn
&+\Delta^A(\Pc) \Delta_B^{rr}(\Qc) \Delta^R(\Kc\Pc) \Delta^A(\Kc\Pc) \Delta_F^{rr}(\Kc\Pc\Qc) \nn
&+\Delta_F^{rr}(\Pc) \Delta^A(\Qc) \Delta^R(\Kc\Pc) \Delta^R(\Kc\Pc) \Delta_F^{rr}(\Kc\Pc\Qc) \nn
&+\Delta^A(\Pc) \Delta^A(\Qc) \Delta_F^{rr}(\Kc\Pc) \Delta^R(\Kc\Pc) \Delta_F^{rr}(\Kc\Pc\Qc) \nn
&+\Delta^A(\Pc) \Delta^R(\Qc) \Delta_F^{rr}(\Kc\Pc) \Delta^A(\Kc\Pc) \Delta_F^{rr}(\Kc\Pc\Qc) \biggr \} \,,
\label{eq:melon1}
\end{align}
where the numerator reads
\begin{align}
F^\mathrm{M1}_{\mu\nu} &= \mathrm{Tr} \left[ \big(i V_\mu\big) \slashed{\Pc} \big(i V_\nu\big) (\slashed{\Kc}+\slashed{\Pc}) \big(i V_{\rho}\big)  (\slashed{\Kc}+\slashed{\Pc}+\slashed{\Qc}) \big(i V_{\sigma}\big) (\slashed{\Kc}+\slashed{\Pc}) \right]  g^{\rho \sigma} \nonumber \\
&= \che^4 \mathrm{Tr} \left[ \gamma_\mu \slashed{\Pc} \gamma_\nu (\slashed{\Kc}+\slashed{\Pc}) \gamma^\rho (\slashed{\Kc}+\slashed{\Pc}+\slashed{\Qc}) \gamma_\rho (\slashed{\Kc}+\slashed{\Pc}) \right] \,.
\end{align}

The expression in \eq\eqref{eq:melon1} contains so-called pinch singularities, which are generated when retarded and advanced propagators with the same momentum arguments multiply each other as in $\Delta^R(\Kc\Pc)\Delta^A(\Kc\Pc)$. They arise when a 0-component integration contour gets squeezed between two poles, which approach a point on the contour from opposite sides. In this case, a pole from a retarded propagator and another from an advanced propagator pinch the contour when taking the limit $\eta \to 0$. Fortunately, these ill-defined terms cancel each other when using \eq\eqref{eq:2distrel} to simplify \eq\eqref{eq:melon1}, leaving us with
\begin{equation}
\begin{split}
-i (\Pi^R_\mathrm{M1}&)_{\mu \nu} (\Kc) = - \int_\Pc \int_\Qc F^\mathrm{M1}_{\mu\nu} (\Kc,\Pc,\Qc) \\
\times  \Big\{&\Delta_F^{rr}(\Pc) \Delta^R(\Kc\Pc)^2 \left[\Delta^A(\Qc) \Delta_F^{rr}(\Kc\Pc\Qc) + \Delta_B^{rr}(\Qc) \Delta^R(\Kc\Pc\Qc) \right] \\
&+\Delta^A(\Pc) \Delta_B^{rr}(\Qc) N_F^-(\Kc\Pc) \left[\Delta^R(\Kc\Pc)^2 \Delta^R(\Kc\Pc\Qc) - \Delta^A(\Kc\Pc)^2 \Delta^A(\Kc\Pc\Qc) \right] \\
&+\Delta^A(\Pc) \Delta_F^{rr}(\Kc\Pc\Qc) N_F^-(\Kc\Pc) \left[\Delta^A(\Qc) \Delta^R(\Kc\Pc)^2 - \Delta^R(\Qc) \Delta^A(\Kc\Pc)^2 \right] \Big\} \,.
\end{split}\label{eq:melon2}
\end{equation}
In some terms, an $rr$-propagator has been eliminated, as its $\delta$-function would trivialize the 0-component integral. 

To proceed further, it is convenient to write the terms with the missing $\delta$-function in terms of the difference $\Delta_2^d\equiv(\Delta^R)^2-(\Delta^A)^2$, which resembles the spectral function $\Delta^d$ except the retarded and advanced propagators are squared. The main motivation behind defining $\Delta_2^d$ is that an integral over $\Delta_2^d$ can be carried out with the help of the  result (that holds for sufficiently regular $f$),
\begin{equation}
\int_{-\infty}^\infty \frac{\mathrm{d}p^0}{2\pi} \Delta_2^d(\Pc) f(p^0) = i\int_{-\infty}^\infty \frac{\mathrm{d}p^0}{2\pi} \Delta^d(\Pc) \frac{\mathrm{d}}{\mathrm{d}p^0}\frac{f(p^0)}{2p^0} \,, \label{eq:p0derpres}
\end{equation}
which effectively reduces it to an integral over $\Delta^d$. \eq\eqref{eq:p0derpres} follows directly from the residue theorem upon (i) rewriting the integral on the left-hand side as a contour integral circulating the two poles of the function $1/(P^2)^2$, (ii) picking up the residues from those poles, and then (iii) deforming the contour back into the integral on the right-hand side. Replacing $(\Delta^A)^2$ with $-\Delta_2^d+(\Delta^R)^2$ in \eq\eqref{eq:melon2}, we find
\begin{equation}
\begin{split}
-i (\Pi^R_\mathrm{M1})_{\mu \nu} &(\Kc) = - \int_\Pc \int_\Qc F^\mathrm{M1}_{\mu\nu} (\Kc,\Pc,\Qc) \\
\times  \Big\{&\Delta_F^{rr}(\Pc) \Delta^R(\Kc\Pc)^2 \left[\Delta^A(\Qc) \Delta_F^{rr}(\Kc\Pc\Qc) + \Delta_B^{rr}(\Qc) \Delta^R(\Kc\Pc\Qc) \right] \\
&+\Delta^A(\Pc) \Delta_2^d(\Kc\Pc) N_F^-(\Kc\Pc) \left[\Delta^R(\Qc) \Delta_F^{rr}(\Kc\Pc\Qc) + \Delta_B^{rr}(\Qc) \Delta^A(\Kc\Pc\Qc) \right] \\
&+\Delta^A(\Pc) \Delta^R(\Kc\Pc)^2 \Delta_B^{rr}(\Qc) \Delta_F^{rr}(\Kc\Pc\Qc) \Big\} \,,
\end{split}\label{eq:melon3}
\end{equation}
where on the last line we have used \eq\eqref{eq:2distrel}. After this procedure, \eq\eqref{eq:melon3} is free of pinch singularities and every term contains a $\delta$-function, trivializing the 0-component integrals.

As before, we change variables so that $\Delta^d$, $\Delta_2^d$, and distribution functions depend only on a single loop momentum. Introducing an additional change of variables,
\begin{equation}
(5) \quad \Pc \mapsto -\Kc-\Pc \,, \quad \Qc \mapsto \Pc+\Qc \,,
\end{equation}
we then arrive at
\begin{equation}
\begin{split}
-i (\Pi^R_\mathrm{M1}&)_{\mu \nu} (\Kc) = - \int_\Pc \int_\Qc \Delta^d(\Qc) \\
  \times \Big\{&\Delta^R(\Kc+\Pc)^2 \Delta^R(\Kc+\Pc+\Qc) \Delta^d(\Pc) N_F^-(\Pc) \left[N_B^{ }(\Qc) F^{\mathrm{M1}}_{\mu\nu} + N_F^+(\Qc) F^{\mathrm{M1}(3)}_{\mu\nu} \right] \\
  &+\Delta^R(\Kc+\Pc) \Delta^R(\Pc+\Qc) \Delta_2^d(\Pc) N_F^+(\Pc) \left[N_B^{ }(\Qc) F^{\mathrm{M1}(2)}_{\mu\nu} + N_F^-(\Qc) F^{\mathrm{M1}(5)}_{\mu\nu} \right] \\
  &+\Delta^R(\Kc+\Pc+\Qc) \Delta^A(\Pc+\Qc)^2 \Delta^d(\Pc) N_F^+(\Pc) N_B^{ }(\Qc) F^{\mathrm{M1}(1)}_{\mu\nu} \Big\} \,.
\end{split}\label{eq:melon4}
\end{equation}
In this case, the changes of variables have not left the numerator $F^\mathrm{M1}_{\mu\nu}$ invariant, as indicated by the notation $F^{\mathrm{M1}(i)}_{\mu \nu}$ which denotes that the change of variables $(i)$ has been performed in the numerator.

\subsubsection{Watermelon diagram 2}
For the second watermelon diagram, applying the Feynman rules gives
\begin{equation}
- \int_\Pc \int_\Qc \mathrm{Tr} \left[ \big(i V_\mu\big) S^p(\Pc) \big(i V_{\rho}\big) S^q(\Pc\Qc) \big(i V_{\sigma}\big) S^r(\Pc) \big(i V_\nu\big) S^s(\Kc\Pc) \right]  D^{\rho \sigma}_t(\Qc) \,.
\end{equation}
The $r/a$ assignments are the same as for the first watermelon, but with the replacement $R \leftrightarrow A$ as the diagrams have the opposite directions of momentum and fermion flows:
\begin{equation}
\begin{split}
-i (\Pi^R_\mathrm{M2})_{\mu \nu} (\Kc) &= - \int_\Pc \int_\Qc F^\mathrm{M2}_{\mu\nu} (\Kc,\Pc,\Qc) \\
\times \biggl\{ &\Delta^A(\Pc) \Delta^A(\Pc) \Delta_B^{rr}(\Qc) \Delta_F^{rr}(\Kc\Pc) \Delta^A(\Pc\Qc) \\
&+\Delta_F^{rr}(\Pc) \Delta^A(\Pc) \Delta_B^{rr}(\Qc) \Delta^R(\Kc\Pc) \Delta^A(\Pc\Qc) \\
&+\Delta_F^{rr}(\Pc) \Delta^R(\Pc) \Delta_B^{rr}(\Qc) \Delta^R(\Kc\Pc) \Delta^R(\Pc\Qc) \\
&+\Delta^A(\Pc) \Delta^R(\Pc) \Delta_B^{rr}(\Qc) \Delta^R(\Kc\Pc) \Delta_F^{rr}(\Pc\Qc) \\
&+\Delta^A(\Pc) \Delta^A(\Pc) \Delta^R(\Qc) \Delta_F^{rr}(\Kc\Pc) \Delta_F^{rr}(\Pc\Qc) \\
&+\Delta_F^{rr}(\Pc) \Delta^A(\Pc) \Delta^R(\Qc) \Delta^R(\Kc\Pc) \Delta_F^{rr}(\Pc\Qc) \\
&+\Delta_F^{rr}(\Pc) \Delta^R(\Pc) \Delta^A(\Qc) \Delta^R(\Kc\Pc) \Delta_F^{rr}(\Pc\Qc) \biggr\} \,.
\end{split}\label{eq:melon5}
\end{equation}
The numerator here reads
\begin{align}
F^\mathrm{M2}_{\mu\nu} &= \mathrm{Tr} \left[ \big(i V_\mu\big) \slashed{\Pc} \big(i V_{\rho}\big) (\slashed{\Pc}+\slashed{\Qc}) \big(i V_{\sigma}\big) \slashed{\Pc} \big(i V_\nu\big) (\slashed{\Kc}+\slashed{\Pc}) \right]  g^{\rho \sigma} \nonumber \\
&= \che^4 \mathrm{Tr} \left[ \gamma_\mu \slashed{\Pc} \gamma^\rho (\slashed{\Pc}+\slashed{\Qc}) \gamma_\rho \slashed{\Pc} \gamma_\nu (\slashed{\Kc}+\slashed{\Pc}) \right] \,.
\end{align}

Since we already manipulated the first watermelon diagram, we are familiar with the necessary steps to take next. By cancelling the pinch singularities, introducing the $\Delta_2^d$ propagator, and performing suitable changes of variables, we arrive at 
\begin{equation}
\begin{split}
-i (&\Pi^R_\mathrm{M2})_{\mu \nu} (\Kc) = - \int_\Pc \int_\Qc \Delta^d(\Qc) \\
  \times \Big\{&\Delta^R(\Kc+\Pc)^2 \Delta^R(\Kc+\Pc+\Qc) \Delta^d(\Pc) N_F^+(\Pc) \left[N_B^{ }(\Qc) F^{\mathrm{M2}(2)}_{\mu\nu} + N_F^-(\Qc) F^{\mathrm{M2}(2,3)}_{\mu\nu} \right] \\
  &+\Delta^R(\Kc+\Pc) \Delta^A(\Pc+\Qc) \Delta_2^d(\Pc) N_F^-(\Pc) \left[N_B^{ }(\Qc) F^{\mathrm{M2}}_{\mu\nu} + N_F^+(\Qc) F^{\mathrm{M2}(6)}_{\mu\nu} \right] \\
  &+\Delta^R(\Kc+\Pc+\Qc) \Delta^R(\Pc+\Qc)^2 \Delta^d(\Pc) N_F^-(\Pc) N_B^{ }(\Qc) F^{\mathrm{M2}(1,2)}_{\mu\nu} \Big\} \,,
\end{split}\label{eq:melon6}
\end{equation}
where we have introduced a sixth variable change,
\begin{equation}
(6) \quad \Qc \mapsto -\Pc-\Qc \,.
\end{equation}
The notation $F^{\mathrm{M2}(i,j)}_{\mu\nu}$ in \eq\eqref{eq:melon6} indicates the result obtained when performing the changes $(i)$ and $(j)$ consecutively in the respective order.

\subsubsection{Trace}
The expressions become quite lengthy when one writes the numerator functions out explicitly. Since we are already familiar with the general structure of the computation from the one-loop case, we choose to highlight here only a few important intermediate steps before presenting the final result. To obtain $(\Pi^R_\mathrm{NLO})_\mu^\mu$, we proceed as follows:
\begin{enumerate}
\item Sum the different contributions from Eqs.~\eqref{eq:cat3},~\eqref{eq:melon4} and~\eqref{eq:melon6} according to \eq\eqref{eq:2loopcontribs}.
\item Contract the numerator functions with $g_{\mu\nu}$ and evaluate the Dirac traces, obtaining factors of $D$ (which turn out to be essential).
\item Use the prescription for $\Delta_2^d$ propagators given by \eq\eqref{eq:p0derpres}.
\item Set the loop momenta on shell, $\Pc^2=\Qc^2=0$, which is justified since every term contains the product $\Delta^d(\Pc) \Delta^d(\Qc)$.
\item Expand in powers of the small external momentum $\Kc$ and keep the leading terms in order to obtain the HTL limit.%
\footnote{Expanding in small $\Kc$ before doing the 0-component integrals is justified as in the one-loop case. As a check, we have also verified that expanding after performing the 0-component integrations yields the same result.}
\item Symmetrize the integrand using
$$\int_{\Pc\Qc} f(\Pc,\Qc) = \int_{\Pc\Qc} (f(\Pc,\Qc) + f(-\Pc,\Qc) + f(\Pc,-\Qc) + f(-\Pc,-\Qc))/4$$
discarding terms that would vanish by symmetry upon integration.
\item Symmetrize the integrand by interchanging the labels of the loop-momenta $\Pc$ and $\Qc$ according to $\int_{\Pc\Qc} f(\Pc,\Qc) = \int_{\Pc\Qc} (f(\Pc,\Qc) + f(\Qc,\Pc))/2$.
\end{enumerate}

After performing the above operations, we have obtained%
\footnote{The retarded prescription has been absorbed into $k^0$. In the final result, we must replace $k^0\to k^0+i\eta$.}
\begin{equation}
\begin{split}
(\Pi^R_\mathrm{NLO})_\mu^\mu (K) &= -\che^4 (D-2) \int_\Pc \int_\Qc \Delta^d(\Pc) \Delta^d(\Qc) \\
  \times \bigg\{&\left(2N_B^{ }(\Qc)-N_F^-(\Qc)-N_F^+(\Qc)\right) \\
&\qquad\times\bigg[(D-2)\frac{\mathrm{d}}{\mathrm{d}p^0}\frac{N_F^-(\Pc)+N_F^+(\Pc)}{2p^0}-\left(N_F^-(\Pc)+N_F^+(\Pc)\right)\frac{\Kc^2}{(\Kc\cdot\Pc)^2} \bigg] \\
&+\frac{1}{2}\left(N_F^-(\Pc)-N_F^+(\Pc)\right)\left(N_F^-(\Qc)-N_F^+(\Qc)\right) \\
&\qquad\times\left[\frac{\Kc^2}{(\Kc\cdot\Pc)(\Kc\cdot\Qc)}-\frac{(\Kc^2)^2 \Pc\cdot\Qc}{(\Kc\cdot\Pc)^2(\Kc\cdot\Qc)^2} \right]\bigg\} \,.\label{eq:2looptrace}
\end{split}
\end{equation}
In the first term inside the braces, the two-loop integral has factorized into two one-loop integrals. In the second term, which vanishes at $\mu=0$, the integrals are coupled and such factorization does not occur. It is also worth noting that before the above symmetrization steps, our expression contained factors of $1/(\Pc\cdot\Qc)$. These factors diverge in $d=3$ dimensions when the three-momenta $\mathbf{p}$ and $\mathbf{q}$ become collinear. Such terms are absent in \eq\eqref{eq:2looptrace}, as these divergences cancel upon symmetrization in the HTL limit. 

Next, we perform the 0-component integrals using \eq\eqref{eq:p0intformula} and express the resulting $d$-dimensional spatial integrals in terms of the radial and angular integrals defined in \app\ref{app:integrals}. This leads to
\begin{equation}
\begin{split}
(\Pi^R_\mathrm{NLO})_\mu^\mu (K) &= -\che^4 \, \mathcal{N}^2 \\
&\times \bigg\{\left(2\mathcal{R}_1-\mathcal{R}_2\right) \mathcal{A}_0 \bigg[\frac{1}{2}(d-1)^2(\mathcal{R}_5-\mathcal{R}_3)\mathcal{A}_0 -(d-1)\Kc^2 \mathcal{R}_3 \mathcal{A}_2 \bigg] \\
&\qquad+\frac{1}{2}(d-1)\mathcal{R}_4^2 \left[\Kc^2\mathcal{A}_1^2+(\Kc^2)^2 \left(\mathcal{A}_2^2-\mathcal{A}_2^i \mathcal{A}_2^i\right) \right]\bigg\} \,,\label{eq:2looptrace2}
\end{split}
\end{equation}
where $\mathcal{N}$ is given by \eq\eqref{eq:intnormfactor}. The only divergent integral in \eq\eqref{eq:2looptrace2} is $\mathcal{R}_3$, which contains a $1/\epsilon$ divergence. 

Let us study the $\mathcal{R}_3$-proportional part of the square brackets on the second line of \eq\eqref{eq:2looptrace2} a bit more closely. It becomes
\begin{equation}
-2\mathcal{R}_3\left\{\left(\mathcal{A}_0+\Kc^2 \mathcal{A}_2\right)-\epsilon\left(2\mathcal{A}_0+\Kc^2 \mathcal{A}_2\right) + O(\epsilon^2)\right\} \,\label{eq:tracedivergences}
\end{equation} 
upon substituting $d = 3-2\epsilon$. By applying the results for the angular integrals, one can show that $\mathcal{A}_0+\Kc^2\mathcal{A}_2 = O(\epsilon)$. Thus, the part inside the curly brackets of \eq\eqref{eq:tracedivergences} becomes $O(\epsilon)$, cancelling the $1/\epsilon$ divergence from the radial integral, and making the whole expression finite.%
\footnote{This UV divergence stems from the self-energy and vertex-correction subgraphs inside the two-loop expressions. The cancellation occurs since the QED renormalization constants $Z_1$ and $Z_2$, respectively for the vertex correction and the electron wavefunction, satisfy $Z_1=Z_2$.} It was essential to evaluate the Dirac trace in $D$ dimensions to obtain this cancellation as the explicit factors of $\epsilon$ in \eq\eqref{eq:tracedivergences} originate from there.

Consequently, after reinserting the finite result of \eq\eqref{eq:tracedivergences} into \eq\eqref{eq:2looptrace2}, and taking $\epsilon\to 0$ everywhere, we arrive to the final result:
\begin{equation}
\begin{split}
(\Pi^R_\mathrm{NLO})_\mu^\mu (K) =& -\frac{\che^4}{8\pi^2}\left(T^2+\frac{\mu^2}{\pi^2}\right)\left(1+\frac{k^0}{k} \log\frac{k^0+k+i\eta}{k^0-k+i\eta}\right) \\
&-\frac{\che^4}{4\pi^2}\frac{\mu^2}{\pi^2}\left(1-\frac{k_0^2}{k^2}\right)\left(1-\frac{k^0}{2k}\log\frac{k^0+k+i\eta}{k^0-k+i\eta}\right)^2 \,. \label{eq:2looptraceresult}
\end{split}
\end{equation}
Here we have reintroduced the retarded prescription $+i\eta$ for $k^0$.

\subsubsection{00-component}
The remaining task is to repeat the above steps for the 00-component of the self-energy. Following the recipe from above \eq\eqref{eq:2looptrace} and picking out the 00-components from the numerator functions (instead of contracting with $g_{\mu\nu}$) leads to
\begin{equation}
\begin{split}
(\Pi^R_\mathrm{NLO})_{00} (K) &= \che^4 (D-2) \frac{1}{2}\int_\Pc \int_\Qc \Delta^d(\Pc) \Delta^d(\Qc) \\
  \times \bigg\{&\left(2N_B^{ }(\Qc)-N_F^-(\Qc)-N_F^+(\Qc)\right) \\
&\times\bigg[-\left(N_F^-(\Pc)+N_F^+(\Pc)\right)\left(\frac{1}{p_0^2}+\frac{2\Kc^2 k^0 p^0}{(\Kc\cdot\Pc)^3}+\frac{\Kc^2}{(\Kc\cdot\Pc)^2}-\frac{2k_0^2}{(\Kc\cdot\Pc)^2}\right) \\
&\qquad +\left(\frac{\mathrm{d}}{\mathrm{d}p^0}\left(N_F^-(\Pc)+N_F^+(\Pc)\right)\right)\left(\frac{1}{p^0}-\frac{\Kc^2 p^0}{(\Kc\cdot\Pc)^2}+\frac{2k^0}{\Kc\cdot\Pc}\right)\bigg] \\
&+\left(N_F^-(\Pc)-N_F^+(\Pc)\right)\left(N_F^-(\Qc)-N_F^+(\Qc)\right) \\
&\qquad\times\left[\frac{(\Kc^2)^2p^0 q^0}{(\Kc\cdot\Pc)^2(\Kc\cdot\Qc)^2}-\frac{2\Kc^2k^0 p^0 }{(\Kc\cdot\Pc)^2(\Kc\cdot\Qc)} +\frac{k_0^2}{(\Kc\cdot\Pc)(\Kc\cdot\Qc)} \right]\bigg\} \,.\label{eq:2loop00}
\end{split}
\end{equation}
In this case, all of the two-loop integrals have factorized into products of one-loop integrals. In addition, collinear divergences have completely cancelled. At this point, we make a remark on the peculiar behavior of the 00-component at small external momentum $\Kc$. Should one consider contributions from the individual diagrams, the 00-components of the two watermelon diagrams would each behave as $O(1/\Kc)$ for small $\Kc$. These contributions are also proportional to odd powers of $\mu$, but thankfully cancel when summing the diagrams together. The leading term in \eq\eqref{eq:2loop00} is thus $O(K^0)$, as expected within the HTL paradigm.

After performing the 0-component integrals in \eq\eqref{eq:2loop00}, we are left with the following $d$-dimensional spatial integrals:
\begin{equation}
\begin{split}
(\Pi^R_\mathrm{NLO})_{00} (K) &= \che^4 (d-1) \frac{1}{2}\mathcal{N}^2\\
  &\times
  \bigg\{\left(2\mathcal{R}_1-\mathcal{R}_2\right) \mathcal{A}_0 \Big[-\mathcal{R}_3\left(\mathcal{A}_0+2\Kc^2 k^0\mathcal{A}_3+\Kc^2 \mathcal{A}_2 -2k_0^2 \mathcal{A}_2 \right) \\
  &\hphantom{{}\times \bigg\{\left(2\mathcal{R}_1-\mathcal{R}_2\right) \mathcal{A}_0 \Big[}
  +\mathcal{R}_5\left(\mathcal{A}_0-\Kc^2\mathcal{A}_2+2k^0\mathcal{A}_1\right)\Big]\\
  &\hphantom{{}\times \bigg\{}
  +\mathcal{R}_4^2 \left[(\Kc^2)^2\mathcal{A}_2^2-2\Kc^2k^0\mathcal{A}_2 \mathcal{A}_1 +k_0^2 \mathcal{A}_1^2 \right]\bigg\} \,. \label{eq:2loop001}
\end{split}
\end{equation}
Once again, the only divergent integral appearing here is $\mathcal{R}_3$, so we inspect the term proportional to it inside the first square brackets:
\begin{equation}
-\mathcal{R}_3\left(\mathcal{A}_0+2\Kc^2 k^0\mathcal{A}_3+\Kc^2 \mathcal{A}_2 -2k_0^2 \mathcal{A}_2 \right) \,.\label{eq:00divergences}
\end{equation}
By evaluating the angular integrals in \eq\eqref{eq:00divergences}, one finds the expression in the parentheses to be $O(\epsilon)$, cancelling the $1/\epsilon$ divergence coming from $\mathcal{R}_3$. Substituting the finite result of \eq\eqref{eq:00divergences} into \eq\eqref{eq:2loop001} and taking $\epsilon \to 0$ everywhere, gives us the final expression for the 00-component:
\begin{equation}
\label{eq:00twoloopfinal}
\begin{split}
(\Pi^R_\mathrm{NLO})_{00} (K) &= \frac{\che^4}{8\pi^2} \left(T^2+\frac{\mu^2}{\pi^2}\right) \left(1+\frac{k_0^2}{\Kc^2} \right) + \frac{\che^4}{4\pi^2} \frac{\mu^2}{\pi^2} \left(1-\frac{k^0}{2k}\log\frac{k^0+k+i\eta}{k^0-k+i\eta}\right)^2 \,.
\end{split}
\end{equation}

\section{Results}
\label{sec:results}
\noindent
In this section, we summarize and inspect the results of our computation. First, we collect all results in a condensed fashion, the study the propagation of soft photons in detail, and finally extract the explicit $O(\varepsilon)$ terms for $\Pi_\mathrm{LO}, \Pi_\mathrm{Pow}$ and $\Pi_\mathrm{NLO}$ in the zero-temperature limit.

\subsection{Summary of the results}
\label{subsec:resultssum}

Here, we summarize the results obtained in the previous section by expressing the self-energy components in the basis of transverse and longitudinal projectors using \eq\eqref{eq:piTandpiL}. We suppress the label $R$ standing for the retarded prescription from now on.

The leading-order (in the coupling $e$) HTL self-energy components for the photon read
\begin{equation}
\label{eq:piLOmain}
\begin{split}
\Pi^{\rm LO}_{\mathrm{T}} &= \frac{\che^2}{2} \bigg (\frac{T^2}{3} + \frac{\mu^2}{\pi^2} \bigg )\left[\frac{k_0^2}{k^2}+\left(1-\frac{k_0^2}{k^2}\right) \frac{k^0}{2k}\log\frac{k^0+k+i\eta}{k^0-k+i\eta}\right], \\
\Pi^{\rm LO}_{\mathrm{L}} & = \che^2\bigg (\frac{T^2}{3} + \frac{\mu^2}{\pi^2} \bigg ) \left(1-\frac{k_0^2}{k^2}\right)\left[1-\frac{k^0}{2k}\log\frac{k^0+k+i\eta}{k^0-k+i\eta}\right],
\end{split}    
\end{equation}
where $\eta>0$. The corresponding renormalized power corrections (see the discussion in \Sec\ref{sec:powcor}) arising from the low-momentum expansion of the one-loop self-energy are 
\begin{equation}
\label{eq:powcormain}
\begin{split}
\Pi^{\rm Pow}_{\mathrm{T}} & = -\frac{\che^2}{4\pi^2} \frac{2K^2}{3} \bigg \{\log \frac{2\exe^{-\gamE}T}{\overline{\Lambda}} - 1  + \frac{1}{4} + \left (1 - \frac{K^2}{4k^2} \right )\left [1 - \frac{k^0}{2k} \log\frac{k^0+k+i\eta}{k^0-k+i\eta} \right ]\\
& \hspace{2.5cm} - \mathrm{Li}^{(1)}_0(-\exe^{\frac{\mu}{T}}) - \mathrm{Li}^{(1)}_0(-\exe^{-\frac{\mu}{T}}) \biggr \} \,,\\
\Pi^{\rm Pow}_{\mathrm{L}}  & = -\frac{\che^2}{4\pi^2} \frac{2K^2}{3} \bigg \{\log \frac{2\exe^{-\gamE}T}{\overline{\Lambda}} - 1  + \left (1 + \frac{K^2}{2k^2} \right ) \left [1 - \frac{k^0}{2k} \log\frac{k^0+k+i\eta}{k^0-k+i\eta} \right ]\\
& \hspace{2.5cm} - \mathrm{Li}^{(1)}_0(-\exe^{\frac{\mu}{T}}) - \mathrm{Li}^{(1)}_0(-\exe^{-\frac{\mu}{T}}) \biggr \} \,,
\end{split}    
\end{equation}
where $\mathrm{Li}^{(1)}_0(z) = \lim_{s \to 0}\frac{\partial \mathrm{Li}_s(z)}{\partial s}$ and $\mathrm{Li}_s$ is the standard polylogarithm function. Note that the UV divergence in these expressions has been eliminated using the photon wavefunction renormalization counterterm in the $\overline{\mathrm{MS}}$-scheme.

The logarithms appearing in \eq\eqref{eq:powcormain} make the limits of vanishing temperature or chemical potential nontrivial. First, we study the power corrections at vanishing chemical potential, for which we get 
\begin{equation}
\lim_{\mu \to 0} \biggl [\mathrm{Li}^{(1)}_0(-\exe^{\frac{\mu}{T}}) + \mathrm{Li}^{(1)}_0(-\exe^{-\frac{\mu}{T}}) \biggr ] = \log \frac{2}{\pi} \,. 
\end{equation}
Hence, we obtain
\begin{align}
\label{eq:powcorTzeromu}
  \lim_{\mu\to 0}\Pi^{\rm Pow}_{\mathrm{T}}  &= -\frac{\che^2}{4\pi^2} \frac{2K^2}{3} \bigg \{ \log \frac{\pi \exe^{-\gamE}T}{\Lbar} - 1  + \frac{1}{4}
 + \left (1 - \frac{K^2}{4k^2} \right )\left (1 - \frac{k^0}{2k} \log\frac{k^0+k+i\eta}{k^0-k+i\eta} \right )\biggr \} \,,
  \\[2mm]
\label{eq:powcorLzeromu}
  \lim_{\mu\to 0}\Pi^{\rm Pow}_{\mathrm{L}} &= -\frac{\che^2}{4\pi^2} \frac{2K^2}{3} \bigg \{\log \frac{\pi \exe^{-\gamE}T}{\Lbar} - 1  + \left (1 + \frac{K^2}{2k^2} \right ) \left (1 - \frac{k^0}{2k} \log\frac{k^0+k+i\eta}{k^0-k+i\eta} \right ) \biggr \} \,.
\end{align}

In the case of vanishing temperature, we can study the leading asymptotic behavior of 
\begin{equation}
\lim_{x \to \infty} \biggl [\mathrm{Li}^{(1)}_0(-\exe^{x}) + \mathrm{Li}^{(1)}_0(-\exe^{-x}) \biggr ] \to -\log x - \gamE \,, 
\end{equation}
where the ratio $x \equiv \mu/T \to \infty$. Hence, we obtain
\begin{align}
\label{eq:powcorTzeroT}
  \lim_{T\to 0}\Pi^{\rm Pow}_{\mathrm{T}} &= -\frac{\che^2}{4\pi^2} \frac{2K^2}{3} \bigg \{\log \frac{2\mu}{\Lbar} - 1  + \frac{1}{4} + \left (1 - \frac{K^2}{4k^2} \right )\left (1 - \frac{k^0}{2k} \log\frac{k^0+k+i\eta}{k^0-k+i\eta} \right )\biggr \} \,,
  \\[2mm]
\label{eq:powcorLzeroT}
  \lim_{T\to 0}\Pi^{\rm Pow}_{\mathrm{L}} &= -\frac{\che^2}{4\pi^2} \frac{2K^2}{3} \bigg \{\log \frac{2\mu}{\Lbar} - 1 + \left (1 + \frac{K^2}{2k^2} \right ) \left (1 - \frac{k^0}{2k} \log\frac{k^0+k+i\eta}{k^0-k+i\eta} \right ) \biggr \} \,.
\end{align}
The expressions obtained in \eqs\nr{eq:powcorTzeromu}--\nr{eq:powcorLzeroT} are in agreement with \Ref\cite{Carignano:2017ovz}.%
\footnote{Note that in \Ref\cite{Carignano:2017ovz} the power corrections are computed in a different regularization scheme.}

Finally, the NLO (in the coupling $e$) contribution to the photon HTL self-energy at nonzero $T$ and $\mu$ reads
\begin{equation}
\label{eq:finalNLOresults}
\begin{split}    
\Pi^{\rm NLO}_{\mathrm{T}} = & - \frac{\che^4}{8\pi^2}\left(T^2+\frac{\mu^2}{\pi^2}\right)\frac{k^0}{2k} \log\frac{k^0+k+i\eta}{k^0-k+i\eta} \,, \\
\Pi^{\rm NLO}_{\mathrm{L}} = & -\frac{\che^4}{8\pi^2} \left(T^2+\frac{\mu^2}{\pi^2}\right) - \frac{\che^4}{4\pi^2} \frac{\mu^2}{\pi^2} \left(1-\frac{k_0^2}{k^2}\right) \left [1-\frac{k^0}{2k}\log\frac{k^0+k+i\eta}{k^0-k+i\eta}\right ]^2 \,.
\end{split}
\end{equation}
These expressions generalize the result obtained in \Ref\cite{Carignano:2019ofj} to nonzero density. As in the zero-$\mu$ case, we find that the intermediate UV and IR singularities are fully canceled, and that the resulting final expression for the NLO contribution is finite.

The terms proportional to $\bigl(T^2 + \frac{\mu^2}{\pi^2}\bigr)$ in $\Pi_\mathrm{T}$ and $\Pi_\mathrm{L}$ follow from the  radial integral 
\begin{equation}
\label{eq:intres1}
\int_0^{\infty} \ud p  p \Bigl (2n_B(p) +n_F(p-\mu) + n_F(p+\mu) \Bigr ) = \frac{\pi^2}{2} \left (T^2 + \frac{\mu^2}{\pi^2}\right ), 
\end{equation}
where the bosonic and fermionic distribution functions have the usual forms $n_{B/F}(x) = (\exe^{x/T} \mp 1)^{-1}$. Note that the fermionic part in \eq\nr{eq:intres1} is identical to the one appearing in the LO expressions in \eq\nr{eq:piLOmain}. In addition, the longitudinal component in \eq\nr{eq:finalNLOresults} includes a new HTL structure containing a squared logarithm, which is only present at finite density. This term arises from the radial integral 
\begin{equation}
\int_0^{\infty} \ud p \Bigl (n_F(p-\mu)  - n_F(p+\mu) \Bigr ) = T\log \left (1 + \exe^{\mu/T} \right ) - T\log \left (1 + \exe^{-\mu/T} \right ) = \mu.   
\end{equation}
Interestingly, at $d = 3$ this integral is independent of $T$. We also observe that due to the more complicated radial integral structure at NLO, the medium dependent mass scale $\mE|_{d=3}$ (defined in \eq\nr{eq:Dmassd3}) does not factorize out from \eq\nr{eq:finalNLOresults} as in the LO case.

We can now apply our results to compute the electric screening length, which follows from $\Pi^{00} = -(k^2/K^2)\Pi_{\rm L}$ in the static infrared limit $\lim_{k \to 0}\Pi^{00}(k_0 = 0, k)$. Utilizing our LO and NLO expressions for $\Pi_{\rm L}$ in \eqs\nr{eq:piLOmain} and \nr{eq:finalNLOresults}, respectively, we obtain 
\begin{equation}
\Pi^{00}(k_0 = 0, \kt \rightarrow 0) = -\biggl [\frac{\che^2}{3} - \frac{\che^4}{8\pi^2} \biggr ]\left (T^2 + \frac{3\mu^2}{\pi^2}\right ) + O(\che^5) \,.
\end{equation}
Note that in QED this quantity is directly related to the equation of state $p(T,\mu)$ via $\che^2\frac{\partial^2p(T,\mu)}{\partial \mu^2} $ ($\che^2 \times$ electric susceptibility). It is straightforward to check that our result is in agreement with known results \cite{Kapusta:2006pm}. We also note that the $O(\che^4)$ result is sensitive to both terms in \eq\nr{eq:finalNLOresults}, including the second one which is only present for $\mu > 0$. This thus constitutes a nontrivial check of our new results.

\subsection{Soft photon propagation at NLO}
\label{subsec:photonmodes}

As an application of our self-energy results, we will next study the NLO corrections to the HTL-resummed soft photon propagator. Let us first recall the main features of the LO case (see e.g.~\Ref\cite{Ghiglieri:2020dpq} for a review). In the time-like region, both transverse and longitudinal components exhibit zero-width plasmon poles  (collective quasiparticle excitations) at the scale $\mE$. At $k=0$ the frequencies of the transverse and longitudinal modes reduce to the plasma frequency, $\omega_\mathrm{p} = \omega_\mathrm{T/L}(k=0) = \pm\mE/\sqrt{3}$, while at high momenta, $k \gg \mE$, the transverse mode acquires an asymptotic mass, $m_\infty = \mE/\sqrt{2}$, and the pole of the longitudinal mode approaches the light cone with an exponentially vanishing residue. On the other hand, in the space-like region the components of the propagator have a Landau cut (originating from the branch cut of the logarithm in the LO self-energies) leading to a non-vanishing spectral function. Next, we consider the NLO corrections to this LO soft photon propagator. Note that in this subsection, we use $\mE$ to denote the three-dimensional in-medium effective mass scale $\mE^2 = \che^2(T^2/3+\mu^2/\pi^2)$ instead of the $d$-dimensional version given by \eq\eqref{eq:g1looptrace3}.

The (dressed) retarded photon propagator is defined in the $\xi$-covariant gauge as
\begin{equation}
\label{eq:DRfull}
\begin{split}
D_R^{\mu\nu}(K) = \prjoT^{\mu\nu}(K)D_\mathrm{T}(K) + \prjoL^{\mu\nu}(K)D_\mathrm{L}(K) - i\xi \frac{K^\mu K^\nu}{K^4} \,,
\end{split}
\end{equation}
where the transverse and longitudinal components read
\begin{equation}\label{eq:propcomps}
D_\mathrm{I}(K) = \frac{-i}{K^2 + \Pi_\text{I}(K)}, \quad  \text{I} \in \{\text{T},\text{L}\} \,,
\end{equation}
and the projectors $\mathcal{P}_\mathrm{I}^{\mu\nu}$ are defined in \Sec\ref{sec:tensorrep}. The retarded prescription further implies that the 0-component of $K$ has a small imaginary part, i.e.~that we need to substitute $k^0\to k^0 +i\eta$, where $\eta>0$. 

It is convenient to study the photon propagator through the spectral function, $\rho^{\mu\nu} \equiv D_R^{\mu\nu} - D_A^{\mu\nu} = 2 \Re D_R^{\mu\nu}$. Moreover, we can inspect the transverse and longitudinal components in a gauge-independent way by projecting $\rho^{\mu\nu}$ onto the conserved currents $J_\mu$ satisfying $K^\mu J_\mu=0$. With $J_\mathrm{L}^i = k^i k^j/k^2 J_j$ being the spatially longitudinal current, we may write
\begin{equation}
J^\mu \rho_{\mu\nu}(K) J^\nu = \left(\mathbf{J}^2-\mathbf{J}^2_\mathrm{L}\right) \rho_\mathrm{T}(K) + \mathbf{J}^2_\mathrm{L} \rho_\mathrm{L}(K) \,,
\end{equation}
where the components of the spectral function are defined by $\rho_\mathrm{I} = 2 \Im \Delta_\mathrm{I}$ with
\begin{equation}
\begin{split}
\Delta_\mathrm{T}(k^0,k) &\equiv \frac{1}{K^2+\Pi_\mathrm{T}(k^0,k)} \,, \\
\Delta_\mathrm{L}(k^0,k) &\equiv -\frac{K^2}{k_0^2} \frac{1}{K^2+\Pi_\mathrm{L}(k^0,k)} \,.
\end{split}
\end{equation}
In our case, the features of the spectral function are qualitatively different in the time-like and space-like regions, so it is useful to consider those regions separately. By assuming that the widths of the plasmon poles are infinitesimal%
\footnote{In QED, the widths of the plasmon poles are beyond-NLO effects.}
and positive in the time-like region, i.e. $\Im(\Delta_\mathrm{I}^{-1}\vert_{k^0\to k^0+i\eta}) = O(\eta)$ and $\mathrm{sgn}\Im(\Delta_\mathrm{I}^{-1}\vert_{k^0\to k^0+i\eta}) = -\mathrm{sgn}(k^0)$ for $\vert k^0\vert >k$, we straightforwardly obtain
\begin{equation}\label{eq:specfunc}
\rho_\mathrm{I}(K) = \vert Z_\mathrm{I}(k)\vert \times 2\pi \, \mathrm{sgn}(k^0) \delta\bigl(k_0^2-\omega_\mathrm{I}^2(k)\bigr)+\rho_\mathrm{I}(K)\theta(k-\vert k^0\vert) \,.
\end{equation}
Here, $\omega_\mathrm{I}(k)$ are the locations of the plasmon poles (dispersion relations), given as the solutions to the equations $\Delta_\mathrm{I}^{-1}(\omega_\mathrm{I},k)=0$, and $Z_\mathrm{I}(k)$ are the corresponding residues, given by
\begin{equation}\label{eq:residuedef}
Z_\mathrm{I}(k) = \frac{1}{1-\Psi_\mathrm{I}(\omega_\mathrm{I},k)} \,,
\end{equation}
where the auxiliary functions $\Psi_\mathrm{I}$ are defined as
\begin{equation}
\begin{split}
\Psi_\mathrm{T}(\omega,k) &\equiv \partial_{\omega^2} \Pi_\mathrm{T}(\omega,k) \,, \\
\Psi_\mathrm{L}(\omega,k) &\equiv \partial_{\omega^2} \frac{\omega^2 \Pi_\mathrm{L}(\omega,k)}{\omega^2-k^2} \,.
\end{split}
\end{equation}
The form of the spectral function in \eq\eqref{eq:specfunc} holds to all loop orders assuming the above conditions hold for the self-energy. 

\begin{figure}[t!]
    \centering
    \includegraphics[scale=0.6]{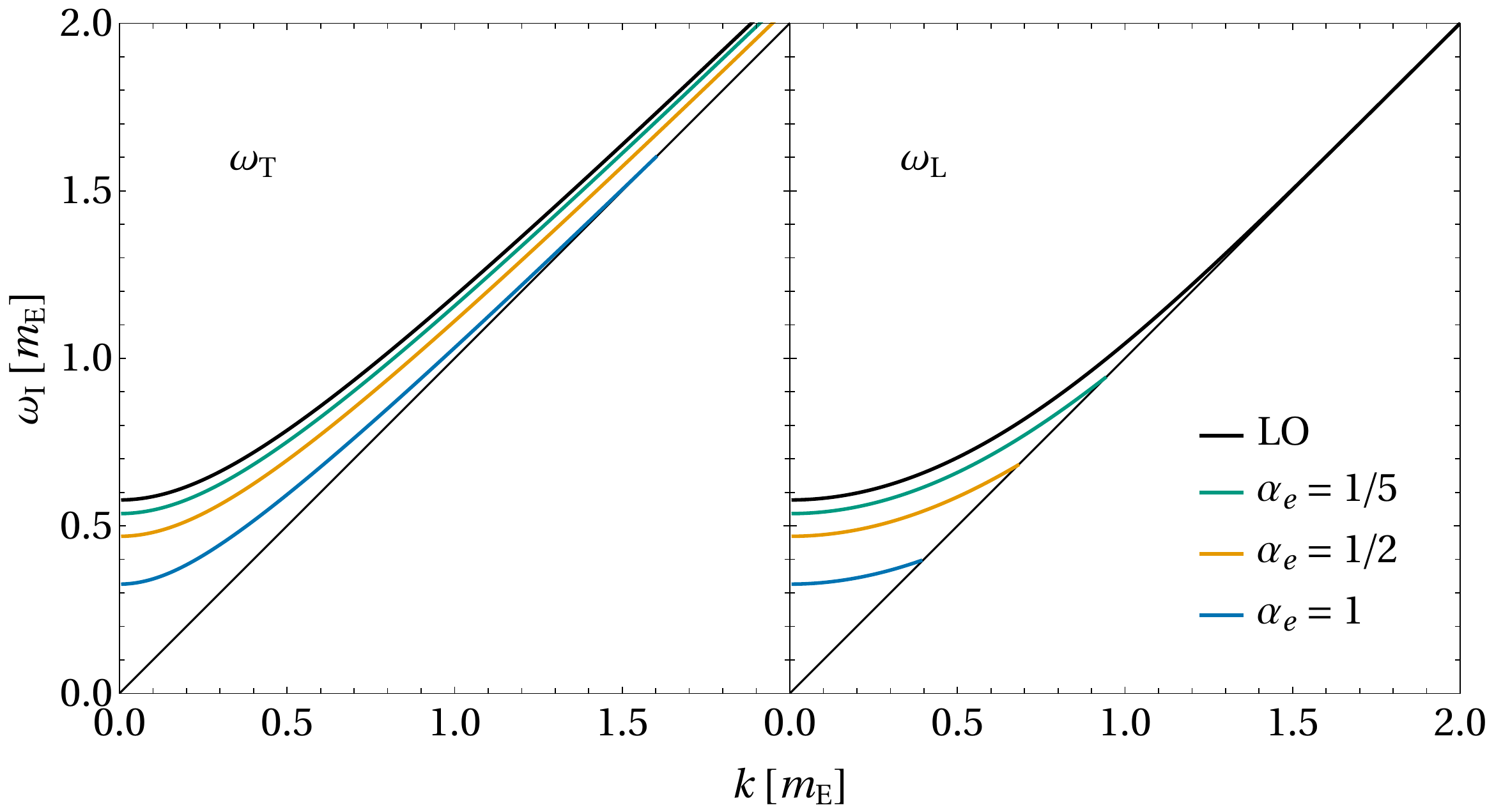}
    \caption{Transverse and longitudinal NLO plasmon dispersion relations at different coupling strengths $\alpha_e$ compared to the LO relations. The values of $\alpha_e$ are chosen rather large for illustrative purposes. For $\alpha_e \lesssim 1$, the results remain qualitatively the same, and the large $\alpha_e$ values are used merely to emphasize the effects of the corrections.}
    \label{fig:disprels}
\end{figure}

We now focus on the NLO corrections to the soft spectral function by using the NLO soft photon self-energy results calculated in this work. First, we consider the locations of the plasmon poles in the time-like region of the spectral function. For soft photons at NLO, the dispersion relation reads
\begin{equation}\label{eq:dispeq}
\begin{split}
  \omega_\mathrm{I}^2 = k^2 + \Pi_\mathrm{I}^\mathrm{LO}(\omega_\mathrm{I}^{ },k) + \delta\Pi_\mathrm{I}^{ }(\omega_\mathrm{I}^{ },k) \,,
\end{split}
\end{equation}
where $\delta\Pi_\mathrm{I}^{ } \equiv \Pi_\mathrm{I}^\mathrm{Pow} + \Pi_\mathrm{I}^\mathrm{NLO}$. To work consistently at order $\che^2 \mE^2$, we expand \eq\nr{eq:dispeq} around the leading order result $\omega_\mathrm{I,LO}$, given by $\omega_\mathrm{I,LO}^2 = k^2 + \Pi_\text{I}^\mathrm{LO}(\omega_\mathrm{I,LO},k)$, as
\begin{equation}
  \omega_\mathrm{I}^2 = \omega_\mathrm{I,LO}^2 + \delta\Pi_\mathrm{I}^{ }(\omega_\mathrm{I,LO},k) + \left(\omega_\mathrm{I}^2-\omega_\mathrm{I,LO}^2\right) \partial_{\omega_\mathrm{I,LO}^2} \Pi_\text{I}^\mathrm{LO}(\omega_\mathrm{I,LO},k) + O(\che^3 \mE^2) \,,
\end{equation}
from which we obtain
\begin{equation}\label{eq:deltaomega}
\delta\omega_\mathrm{I}^2 \equiv \omega_\mathrm{I}^2-\omega_\mathrm{I,LO}^2 = \frac{\delta\Pi_\mathrm{I}(\omega_\mathrm{I,LO},k)}{1-\partial_{\omega_\mathrm{I,LO}^2} \Pi_\text{I}^\mathrm{LO}(\omega_\mathrm{I,LO},k)} + O(\che^3 \mE^2) \,.
\end{equation}

\begin{figure}[t!]
    \centering
    \includegraphics[width=\textwidth]{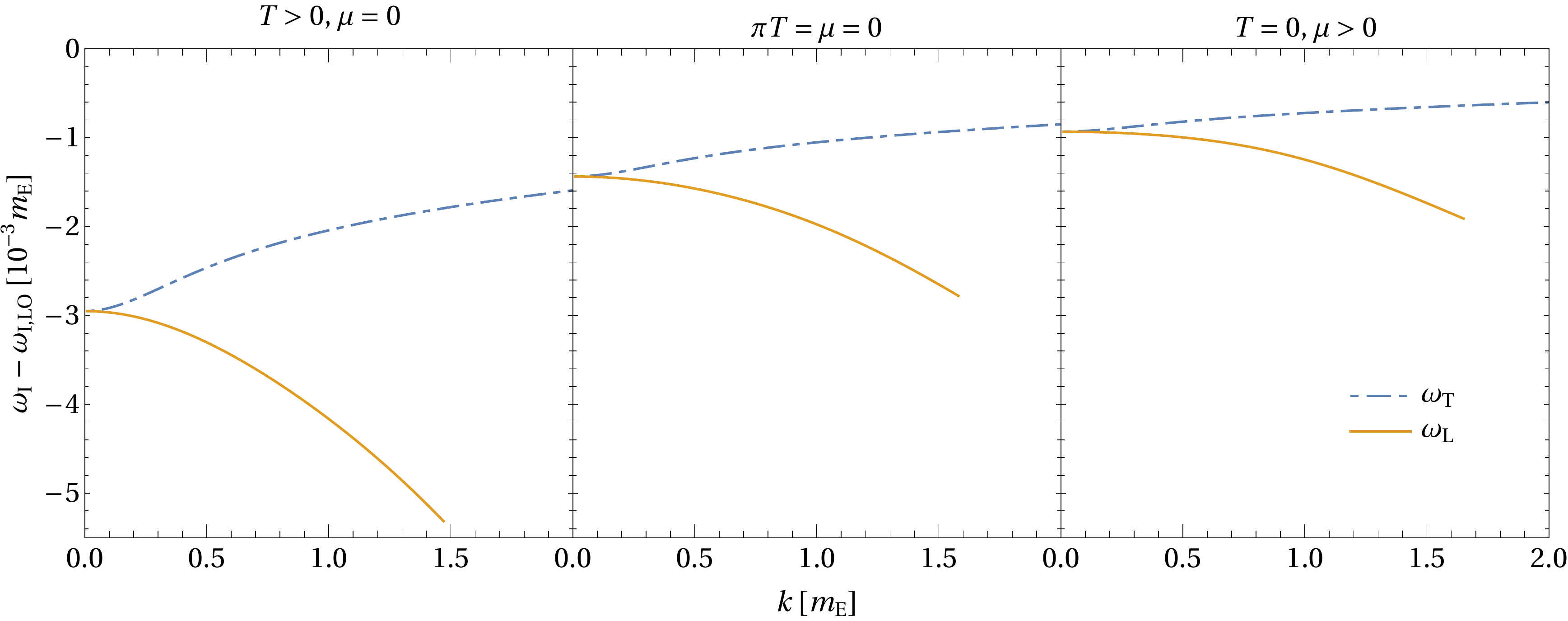}
    \caption{Difference between the NLO and LO dispersion relations evaluated at various $\mu/T$. The longitudinal component has been plotted only up to the point where it crosses the lightcone. Here $\alpha_e = 1/137$.}
    \label{fig:disprelsdiff}
\end{figure}

Applying the LO self-energy results from \eq\nr{eq:piLOmain} to \eq\nr{eq:deltaomega} then yields
\begin{equation}
\begin{split}
\omega_\mathrm{T}^2 &= \omega_\mathrm{T,LO}^2 + \frac{2\omega_\mathrm{T,LO}^2(k^2-\omega_\mathrm{T,LO}^2)}{(k^2-\omega_\mathrm{T,LO}^2)^2-\mE^2\omega_\mathrm{T,LO}^2} \delta\Pi_\mathrm{T}(\omega_\mathrm{T,LO},k) \,, \\
\omega_\mathrm{L}^2 &= \omega_\mathrm{L,LO}^2 + \frac{2\omega_\mathrm{L,LO}^2}{k^2-\omega_\mathrm{L,LO}^2+\mE^2} \delta\Pi_\mathrm{L}(\omega_\mathrm{L,LO},k) \,,
\end{split}
\end{equation}
where we dropped the higher order $O(\che^3 \mE^2)$ terms. These NLO dispersion relations are shown in \fig\ref{fig:disprels} with different values of the fine-structure constant $\alpha_e = e^2/4\pi$.%
\footnote{For convenience, the renormalization scale $\Lbar$ is chosen such that $\log \left(2\exe^{-\gamE}T/\overline{\Lambda}\right) - 1 - \mathrm{Li}^{(1)}_0(-\exe^{\frac{\mu}{T}}) - \mathrm{Li}^{(1)}_0(-\exe^{-\frac{\mu}{T}}) = 0$. Further, we choose $\pi T = \mu > 0$. These apply to all figures unless otherwise stated.}
The effect of varying $\mu/T$ is further illustrated in \fig\ref{fig:disprelsdiff} by plotting the difference between the NLO and LO results. As indicated by the figures, the NLO correction shifts the LO result downward and causes the dispersion curve to pierce the light cone at a finite value of $k$. For the transverse component, the piercing happens only at very high values of $k/\mE$ $(\sim \exp(1/\che^2))$ unless $\alpha_e$ is large [i.e.~$O(1)$], but for the longitudinal component, the dispersion curve hits the light cone at relatively small values of $k/\mE$ ($\sim \log(1/\che^{1/2})$) even with small $\alpha_e$.

\begin{figure}[t!]
    \centering
    \includegraphics[scale=0.6]{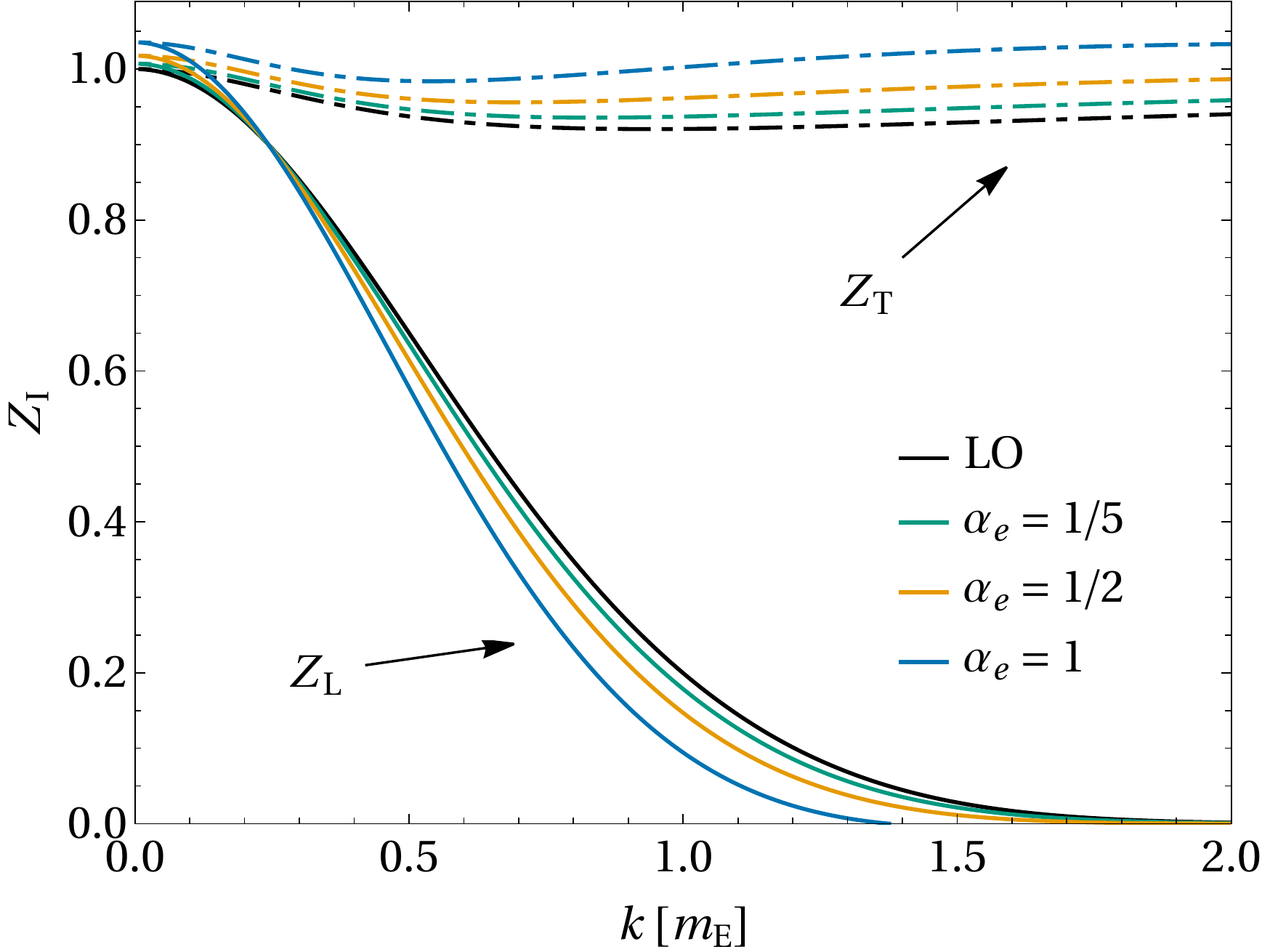}
    \caption{Transverse and longitudinal NLO residues of the plasmon poles at various values of $\alpha_e$ compared to the LO case.}
    \label{fig:residues}
\end{figure}

Next, we study the residues of the transverse and longitudinal plasmon poles at NLO. Expanding the all-order result in \eq\nr{eq:residuedef} around the LO one, given by $Z_\mathrm{I}^\mathrm{LO} = 1/[1-\Psi_\mathrm{I}^\mathrm{LO}(\omega_\mathrm{I,LO},k)]$, results in
\begin{equation}
  Z_\mathrm{I}^{ } = Z_\mathrm{I}^\mathrm{LO} \left\{1+Z_\mathrm{I}^\mathrm{LO}\left[\delta\Psi_\mathrm{I}(\omega_\mathrm{I,LO},k)+\delta\omega_\mathrm{I}^2 \partial_{\omega_\mathrm{I,LO}^2}\Psi_\mathrm{I}^\mathrm{LO}(\omega_\mathrm{I,LO},k)\right]\right\} + O(\che^3) \,,
\end{equation}
where the $O(\che^3)$ terms may be dropped as higher-order contributions. In \fig\ref{fig:residues}, we plot the above NLO residues of the transverse and longitudinal photon modes. As seen in the figure, the correction to the transverse residue is positive for the values of momenta $k$ shown, whereas the one for the longitudinal component starts positive and eventually changes sign as $k$ increases. While the LO result for the longitudinal residue approaches zero exponentially, it is worth noting that the corrected results reach zero at a finite value of $k$.%
\footnote{We have observed that the NLO longitudinal residue goes to zero only if $T>0$.}

\begin{figure}[t!]
    \centering
    \includegraphics[scale=0.6]{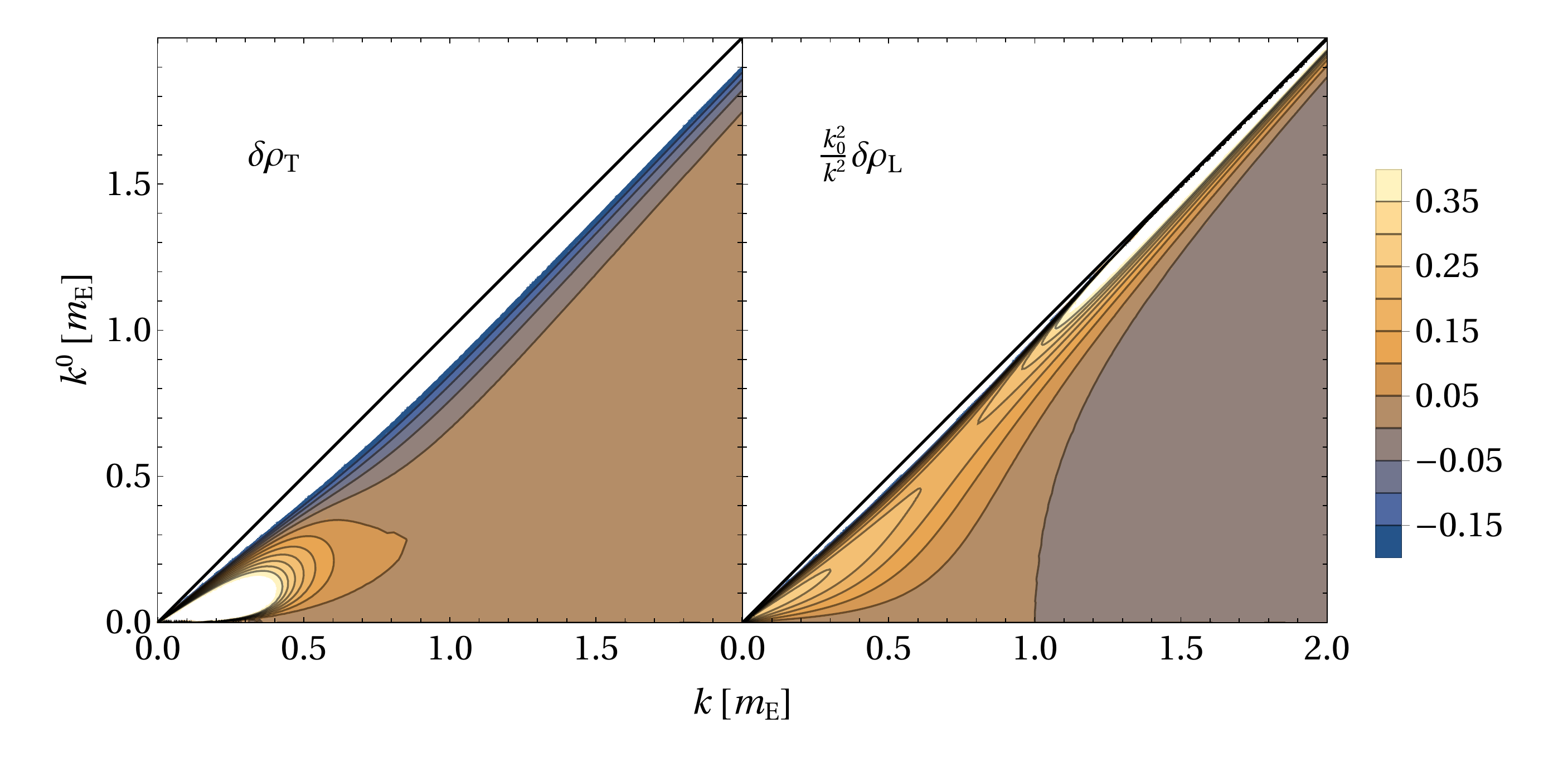}
    \caption{NLO corrections to the transverse and longitudinal components of the spectral function in the Landau cut. Note that the longitudinal component has been scaled by $k_0^2/k^2$ here as well as in \figs\ref{fig:cut} and \ref{fig:cutfixk}. Here $\alpha_e=1/2$.}
    \label{fig:cutdelta}
\end{figure}

\begin{figure}
    \centering
    \includegraphics[scale=0.6]{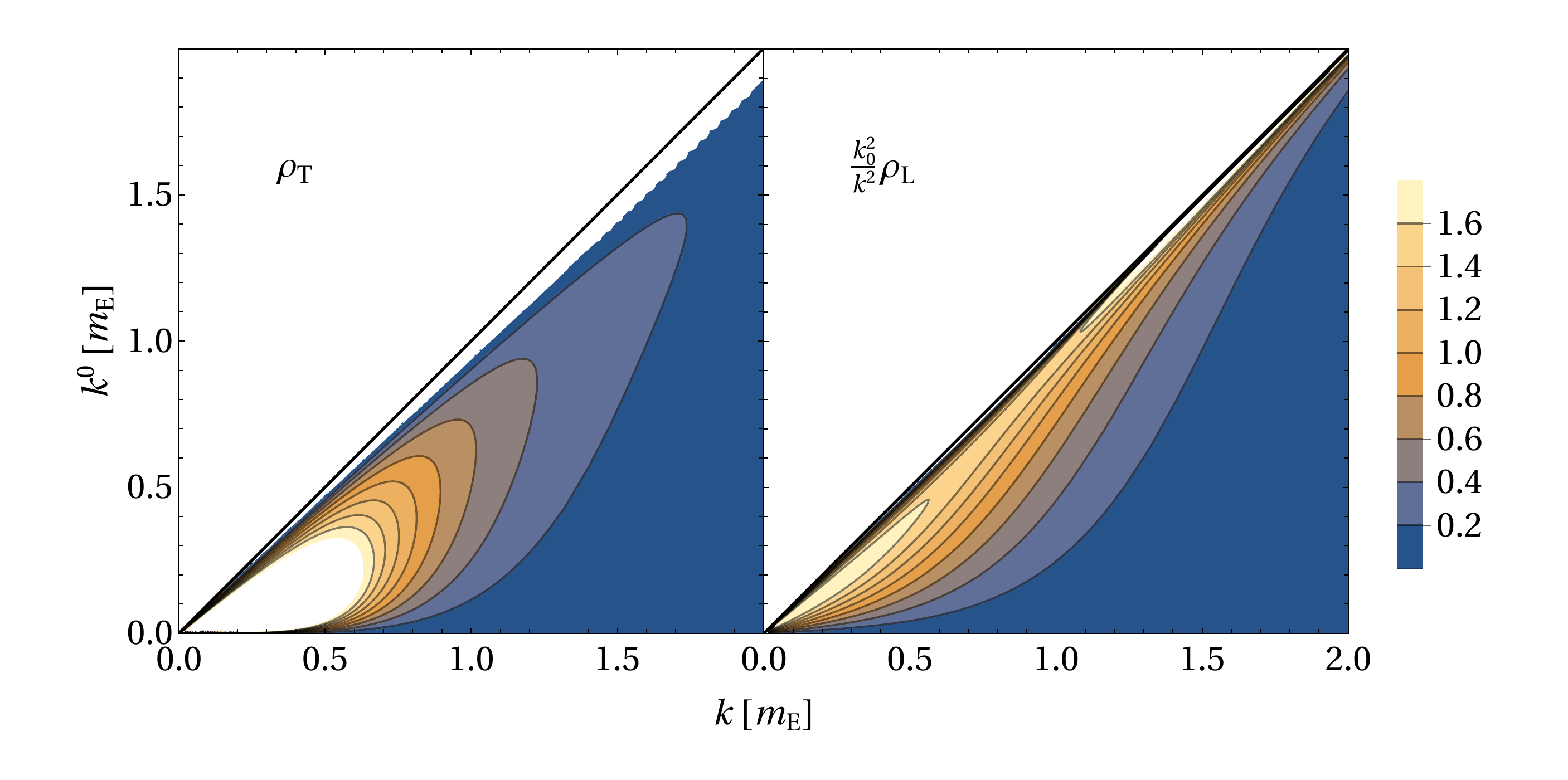}
    \caption{Transverse and longitudinal components of the NLO photon spectral function with $\alpha_e=1/2$. Vertical slices at $k=\mE/2$ are shown in \fig\ref{fig:cutfixk}.}
    \label{fig:cut}
\end{figure}

\begin{figure}[t!]
    \centering
    \includegraphics[scale=0.6]{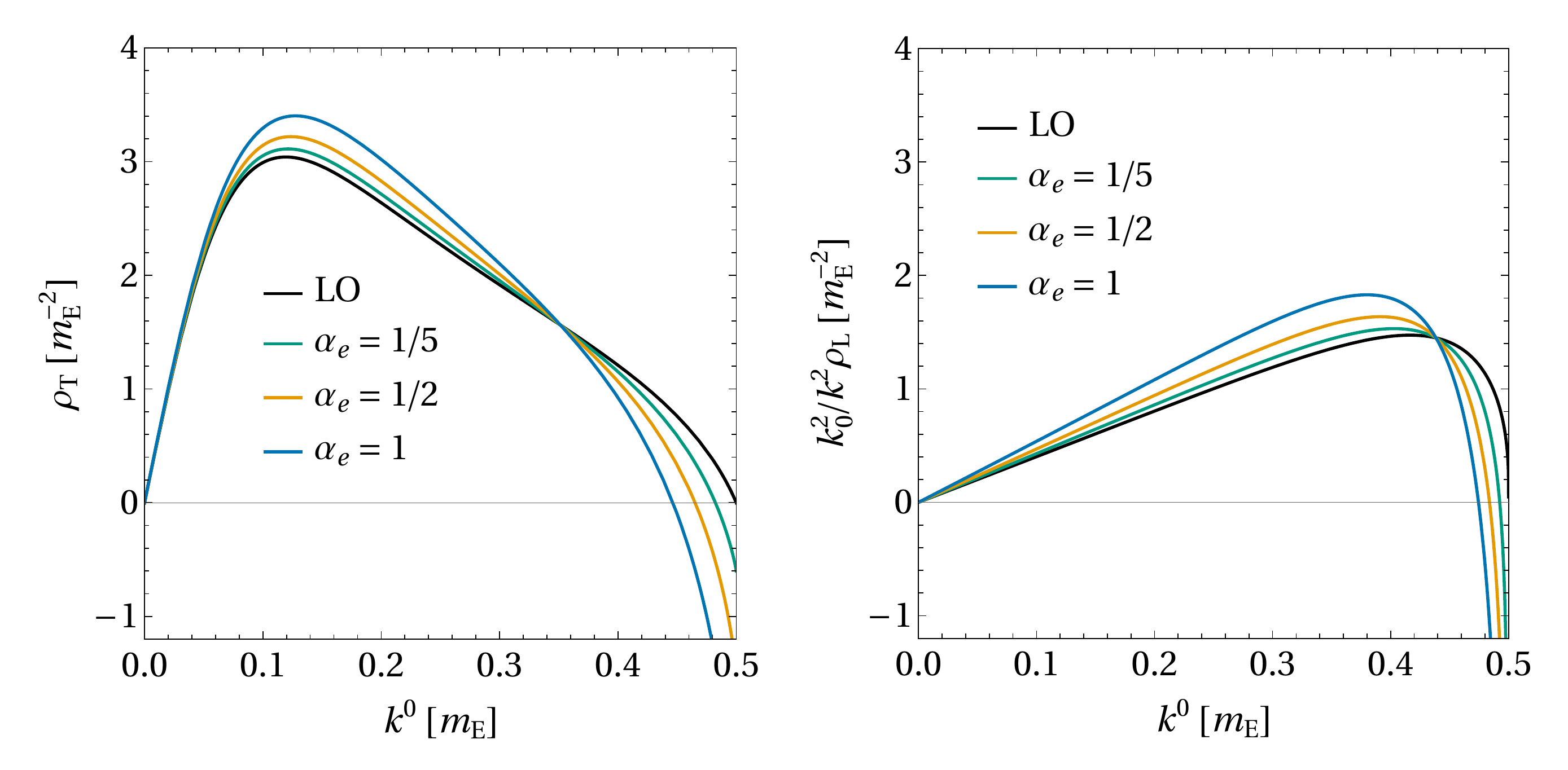}
    \caption{NLO photon spectral functions with fixed momentum $k=\mE/2$ compared to the LO ones.}
    \label{fig:cutfixk}
\end{figure}

Finally, we consider the spectral function in \eq\eqref{eq:specfunc} in the space-like region, $k > \vert k^0\vert$, where it remains non-zero due to the Landau cut in the propagator. Expanding around the leading order results $\rho_\mathrm{I}^\mathrm{LO}=2\Im\Delta_\mathrm{I}^\mathrm{LO}$ leads to the NLO corrections
\begin{equation}
  \delta\rho_\mathrm{I}^{ } \equiv \rho_\mathrm{I}^{ } - \rho_\mathrm{I}^\mathrm{LO} = -2 \Im \biggl(\Delta_\mathrm{I}^\mathrm{LO} \frac{\delta \Pi_\mathrm{I}}{K^2+\Pi_\mathrm{I}^\mathrm{LO}}\biggr) + O(\che^3 \mE^{-2}) \,,
\end{equation}
where we may drop the higher order $O(\che^3\mE^{-2})$ terms. These corrections to the components of the soft spectral function in the Landau cut are plotted for $\alpha_e=1/2$ in \fig\ref{fig:cutdelta} alongside the full NLO results in \fig\ref{fig:cut}. Further, to visualize the corrections at various coupling strengths, we plot slices of the spectral function with fixed momentum in \fig\ref{fig:cutfixk}. We observe from the figures that for $\pi T=\mu>0$ the NLO corrections are mainly positive except in the vicinity of the light cone, where even the full NLO spectral function becomes negative. The near-light cone behavior stays qualitatively the same with different values of $\mu/T$.

The above results for the NLO HTL-resummed photon propagator (spectral function) display peculiar behavior near the light cone. The plasmon dispersion curves hitting the light cone together with the negative spectral function in the Landau cut is related to a breakdown of the HTL-resummation as $K^2\to 0$. Indeed, the NLO correction of relative order $\che^2$ eventually becomes greater than the LO result as $K^2\to 0$, which can be seen by noting that $\Pi_\mathrm{NLO}/K^2$  diverges faster than $\Pi_\mathrm{LO}/K^2$ as $K^2 \to 0$ (cf.~\eqs\eqref{eq:finalNLOresults} and \eqref{eq:piLOmain}). 
This breakdown of perturbation theory implies that we can  trust our results only up to some small distance from the light cone.  It also (partly) explains the unexpected behavior of the resummed propagator as $K^2\to 0$. However, the longitudinal dispersion curve passing through the light cone is a physical effect that has already been predicted, at least in
scalar QED~\cite{Kraemmer:1994az},
non-relativistic QED~\cite{Tsytovich61}, and even in
QCD~\cite{Silin:1988js,Lebedev:1989ev}.

In scalar QED, the failure of the NLO HTL resummation near the light cone has been investigated thoroughly in \Ref\cite{Kraemmer:1994az}.  There, the NLO soft photon dispersion relations display qualitatively similar behavior to the presently studied case of physical QED, even though the NLO corrections in scalar QED arise in an entirely different way.%
\footnote{In scalar QED, the NLO corrections to the soft photon self-energy arise from fully soft HTL-resummed one-loop diagrams resulting in $O(\che\mE^2)$ terms  (cf.~the $\che^2\mE^2$ corrections presented in this paper).}
In Ref.~\cite{Kraemmer:1994az}, the breakdown of the HTL resummation is caused by hard massless particles running in the loops, which leads to collinear singularities in the small-$K$ expansion of $\Pi^{\mu\nu}$ at $K^2=0$. We suspect that to obtain correct behavior of the HTL expressions in our case one must also carry out additional resummations, such as including ladder-type diagrams~\cite{Aurenche:2002wq} when considering photon propagation near the light cone.

In the perturbative region sufficiently far from the light cone, our results are applicable for soft momenta, $k\lesssim \mE$. From the obtained NLO results, we can then take various IR limits, provided that $k^0 \neq k$. For instance, the NLO photon plasma frequency $\omega_\mathrm{p}$, given by the equation $\omega_\mathrm{p}^2 = \Pi_\mathrm{T}(\omega_\mathrm{p},0) = \Pi_\mathrm{L}(\omega_\mathrm{p},0)$, reads
\begin{equation}
\omega_\mathrm{p}^2 = \frac{\mE^2}{3}\left\{1-\frac{\che^2}{6\pi^2} \left[\frac{37}{12}+\frac{9\pi^2T^2}{2\pi^2T^2+6\mu^2}
-\log \frac{2\exe^{-\gamE}T}{\overline{\Lambda}} + \mathrm{Li}^{(1)}_0(-\exe^{\frac{\mu}{T}}) + \mathrm{Li}^{(1)}_0(-\exe^{-\frac{\mu}{T}})\right] \right\} \,.
\end{equation}
On the other hand, the NLO Debye mass (electric screening mass), defined here by $\mD^2 = -\Pi_{00}(0,k)|_{k^2\to -\mD^2}$, is given by%
\footnote{It is easy to check that $\omega_\mathrm{p}$ and $\mD$ are renormalization scale invariant up to and including $O(e^2 \mE^2)$ terms, $\mathrm{d}\mD^2/\mathrm{d}\overline{\Lambda} = \mathrm{d}\omega_\mathrm{p}^2/\mathrm{d}\overline{\Lambda} = 0$.}
\begin{equation}
\mD^2 = \mE^2 \left\{1-\frac{\che^2}{6\pi^2}\left[\frac{7}{4}-\log \frac{2\exe^{-\gamE}T}{\overline{\Lambda}} + \mathrm{Li}^{(1)}_0(-\exe^{\frac{\mu}{T}}) + \mathrm{Li}^{(1)}_0(-\exe^{-\frac{\mu}{T}})\right]\right\} \,.
\end{equation}
Obtaining the asymptotic mass of the transverse photon mode $m_\infty$ to NLO, however, requires information about the self-energy near the light cone. As our results become ill-behaved in this region, we cannot extract the full NLO value of this quantity. Nevertheless, we find a contribution to the NLO asymptotic mass of the transverse photon mode, given by $\omega^2_\mathrm{T}(k\gg\mE)=k^2+m_\infty^2$, 
\begin{equation}
m_\infty^2 = \frac{\mE^2}{2} \left[1+\frac{\che^2 \log \che}{12\pi^2} \left( \frac{11\pi^2T^2+15\mu^2}{\pi^2T^2+3\mu^2} \right) +O(e^2)\right] \,,
\end{equation}
where we exclude the relative order $O(e^2)$ terms.
Those terms contain logarithmic $k$-dependence%
\footnote{We find that this logarithmic $k$-dependence still remains even after renormalization of the LO result.}
and we expect them to be related to the breakdown of our expressions near the light cone.

\subsection{\texorpdfstring{The $O(\varepsilon)$ terms in the zero-temperature limit}{The order epsilon terms in the zero-temperature limit}}
\label{coldstuff}

Next, we consider the $O(\varepsilon)$ terms for $\Pi_\mathrm{LO}, \Pi_\mathrm{Pow}$, and $\Pi_\mathrm{NLO}$. These terms become useful when considering higher-order diagrams in which the self-energy appears. Here, we only write the explicit results in the zero-temperature limit, which become relevant e.g.~in the calculation of the cold and dense QED pressure~\cite{Seppanen}. Following the notation introduced in \Ref\cite{Gorda:2021kme}, we first expand the $d$-dimensional $(d = 3-2\varepsilon)$ expressions for the self-energies determined in \Sec\ref{sec:detailedcomputation} in a series 
\begin{equation}\label{eq:piepsexpansion}
(\Pi^R_\mathrm{X})_{\mathrm{I}} = (\Pi^R_\mathrm{X})^{(0)}_{\mathrm{I}} + \varepsilon (\Pi^R_\mathrm{X})^{(1)}_{\mathrm{I}} + O(\varepsilon^2) \,, \quad \rm{I} \in \{\rm T, \rm L\}
\;,
\end{equation}
where the subscript $\rm X \in \{\rm LO, \rm NLO, \rm Pow\}$, and then determine the unknown coefficients. Here, $(\Pi^R_\mathrm{Pow})_{\mathrm{I}}$ also contains a single vacuum UV divergence so the corresponding expansion in \eq\nr{eq:piepsexpansion} also has a  $(\Pi^R_\mathrm{Pow})^{(-1)}_{\mathrm{I}}$ term.

It is straightforward to extract the $O(\varepsilon)$ terms in \eq\nr{eq:piepsexpansion} using the $d$-dimensional results for the radial and angular integrals in \app\ref{app:integrals}. Applying these results to the LO self-energy in \eqs\nr{eq:g1looptrace3} and \nr{eq:g1loop001} yields 
\begin{equation}
\begin{split}
(\Pi^R_\mathrm{LO})_\mu^\mu (K) & =  (\Pi^R_\mathrm{LO})^{(0) \mu}_\mu (K)  + \varepsilon (\Pi^R_\mathrm{LO})^{(1) \mu}_\mu (K) + O(\varepsilon^2) \,, \\
(\Pi^R_\mathrm{LO})_{00} (K) & =  (\Pi^R_\mathrm{LO})^{(0)}_{00}(K)  + \varepsilon (\Pi^R_\mathrm{LO})^{(1)}_{00} (K) + O(\varepsilon^2) \,, 
\end{split}
\end{equation}
where the coefficients $(\Pi^R_\mathrm{LO})^{(i) \mu}_\mu(K)$ and $(\Pi^R_\mathrm{LO})^{(i)}_{00}(K)$  for $i = 0,1$ are given by
\begin{equation}
\begin{split}
(\Pi^R_\mathrm{LO})^{(0) \mu}_\mu (K) & = \mE^2 \,, \\
(\Pi^R_\mathrm{LO})^{(1) \mu}_\mu (K) & = 0 \,,
\end{split}
\end{equation}
and
\begin{equation}
\begin{split}
(\Pi^R_\mathrm{LO})^{(0)}_{00} (K) & = -\mE^2 \left(1-k^0 L(K)\right) \,, \\
(\Pi^R_\mathrm{LO})^{(1)}_{00} (K) & = -\mE^2 k^0 H(K) \,.
\end{split}
\end{equation}
The function $L(K)$ is defined in \eq\nr{eq:HTLlogarithm} and we further define
\begin{equation}
H(K) \equiv L(K) \left[2+\log\left(\frac{K^2}{4k^2}\right)\right] - \frac{1}{2k} \left[\mathrm{Li}_2\left(\frac{k^0+k}{k^0-k}\right)-\mathrm{Li}_2\left(\frac{k^0-k}{k^0+k}\right)\right] \,.
\end{equation}
Note that the trace of the one-loop HTL self-energy in \eq\nr{eq:g1looptrace3} is defined to be $\mE^2$. In $d$ dimensions, the in-medium effective mass scale at nonzero density reads (see also \Ref\cite{Gorda:2021kme})
\begin{equation}
\begin{split}
\mE^2 & = \che^2\mu^2 \left (\frac{\exe^{\gamE}\Lh^2}{4\pi\mu^2} \right )^{\frac{3-d}{2}} \frac{8\Gamma(\frac{1}{2})}{(4 \pi)^\frac{d+1}{2}\Gamma\left(\frac{d}{2}\right)} \\
& = \frac{\che^2\mu^2}{\pi^2} + O(\varepsilon) \,,
\end{split}
\end{equation}
which agrees with \eq\eqref{eq:Dmassd3}. The coefficients of the expanded transverse and longitudinal components are finally given by
\begin{equation}
\begin{split}
(\Pi^R_\mathrm{LO})^{(0)}_{\mathrm{T}} (K) &= \frac{\mE^2}{2} \left(\frac{k_0^2}{k^2}+\left(1-\frac{k_0^2}{k^2}\right)k^0 L(K) \right) \,, \\
(\Pi^R_\mathrm{LO})^{(1)}_{\mathrm{T}} (K) &= \frac{\mE^2}{2} \biggl\{1- \left(1-\frac{k_0^2}{k^2}\right)\left(1-k^0 L(K)+k^0 H(K)\right) \biggr\} \,,
\end{split}
\end{equation}
and
\begin{equation}
\begin{split}
(\Pi^R_\mathrm{LO})^{(0)}_{\mathrm{L}} (K) &= \mE^2 \left(1-\frac{k_0^2}{k^2}\right) \left(1-k^0 L(K)\right) \,, \\
(\Pi^R_\mathrm{LO})^{(1)}_{\mathrm{L}} (K) &= \mE^2\left(1-\frac{k_0^2}{k^2}\right) k^0 H(K) \,.
\end{split}
\end{equation}

At NLO, the expression for the trace part of the HTL self-energy in \eq\nr{eq:2looptrace2}, generalized to $d$ dimensions, reads
\begin{equation}
\begin{split}
(\Pi^R_\mathrm{NLO})_\mu^\mu (K) =  (\Pi^R_\mathrm{NLO})^{(0) \mu}_\mu (K)  + \varepsilon (\Pi^R_\mathrm{NLO})^{(1) \mu}_\mu (K) + O(\varepsilon^2) \,,
\end{split}
\end{equation}
where the coefficients $(\Pi^R_\mathrm{NLO})^{(i) \mu}_\mu(K)$ for $i = 0,1$ are given by
\begin{equation}
\begin{split}
(\Pi^R_\mathrm{NLO})^{(0) \mu}_\mu (K) & = -\frac{\mE^4}{8\mu^2} \bigg\{ 1+2k^0 L(K) + 2\left(1-\frac{k_0^2}{k^2}\right)\left(1-k^0 L(K)\right)^2 \bigg\} \,, \\
(\Pi^R_\mathrm{NLO})^{(1) \mu}_\mu (K) & = \frac{\mE^4}{4\mu^2}\bigg\{1+\left(1-\frac{k_0^2}{k^2}\right) \left(1-k^0 L(K)\right)^2 \\
&\hphantom{{}=\frac{\mE^4}{4\mu^2}\bigg\{1}
  +\left[1-2\left(1-\frac{k_0^2}{k^2}\right) \left(1-k^0 L(K)\right)\right] k^0 H(K) \bigg\} \,.
\end{split}
\end{equation}
Similarly, the NLO expression for the $00$-component in \eq\nr{eq:2loop001}, also generalized to $d$ dimensions, yields
\begin{equation}
\begin{split}
(\Pi^R_\mathrm{NLO})_{00} (K) =  (\Pi^R_\mathrm{NLO})^{(0)}_{00}(K)  + \varepsilon (\Pi^R_\mathrm{NLO})^{(1)}_{00}(K) + O(\varepsilon^2) \,,
\end{split}
\end{equation}
where the coefficients $(\Pi^R_\mathrm{NLO})^{(i)}_{00} (K)$ for $i = 0,1$ are given by
\begin{equation}
\begin{split}
(\Pi^R_\mathrm{NLO})^{(0)}_{00} (K) & = \frac{\mE^4}{8\mu^2} \left\{\frac{k^2}{K^2} + 2\left(1-k^0 L(K)\right)^2 \right\} \,, \\
(\Pi^R_\mathrm{NLO})^{(1)}_{00} (K) & = -\frac{\mE^4}{4\mu^2} \left(1-k^0 L(K)\right)\bigg\{ 1-k^0 L(K) +\frac{k^2}{K^2} -2k^0 H(K) \bigg\} \,.
\end{split}
\end{equation}
The coefficients of the expanded transverse and longitudinal components read
\begin{equation}
\begin{split}
(\Pi^R_\mathrm{NLO})^{(0)}_{\mathrm{T}} (K) &= -\frac{\mE^4}{8\mu^2} k^0 L(K) \,, \\
(\Pi^R_\mathrm{NLO})^{(1)}_{\mathrm{T}} (K) &= \frac{\mE^4}{8\mu^2} k^0 H(K) \,,
\end{split}
\end{equation}
and
\begin{equation}
\begin{split}
(\Pi^R_\mathrm{NLO})^{(0)}_{\mathrm{L}} (K) &= -\frac{\mE^4}{8\mu^2} \left\{1 +2 \left( 1-\frac{k_0^2}{k^2} \right) \left(1-k^0 L(K)\right)^2 \right\} \,, \\
(\Pi^R_\mathrm{NLO})^{(1)}_{\mathrm{L}} (K) &= \frac{\mE^4}{4\mu^2} \left(1-k^0 L(K)\right) \bigg\{ 1+ \left(1-\frac{k_0^2}{k^2}\right) \left( 1-k^0 L(K)- 2k^0 H(K) \right) \bigg\} \,.
\end{split}
\end{equation}

Finally, we generalize the power corrections to the one-loop HTL self-energy in \eq\nr{eq:powcormain} to $d$ dimensions.  The coefficients of the expanded transverse and longitudinal components end up reading
\begin{equation}
\begin{split}
(\Pi^R_{\rm Pow})^{(-1)}_{\mathrm{T}}  & = \frac{\che^2}{4\pi^2} \frac{K^2}{3} \,, \\
(\Pi^R_{\rm Pow})^{(0)}_{\mathrm{T}}  & = -\frac{\che^2}{8\pi^2} \frac{K^2}{3} \left\{1-2I+\left(3+\frac{k_0^2}{k^2}\right)\left(1-k^0 L(K)\right) \right\} \,, \\
(\Pi^R_{\rm Pow})^{(1)}_{\mathrm{T}}  & = \frac{\che^2}{8\pi^2} \frac{K^2}{3} \bigg\{4-\frac{\pi^2}{2}-I+I^2 -\left(3+\frac{k_0^2}{k^2}\right) k^0 H(K)  \\
  &\hphantom{{}=\frac{\che^2}{8\pi^2} \frac{K^2}{3} \bigg\{4}
  +\left[2-4I-(2-I)\left(1-\frac{k_0^2}{k^2}\right) \right]\left(1-k^0 L(K)\right) \bigg\} \,,
\end{split}    
\end{equation}
and
\begin{equation}
\begin{split}
(\Pi^R_{\rm Pow})^{(-1)}_{\mathrm{L}} & = \frac{\che^2}{4\pi^2} \frac{K^2}{3} \,, \\
(\Pi^R_{\rm Pow})^{(0)}_{\mathrm{L}} & = -\frac{\che^2}{4\pi^2} \frac{K^2}{3} \bigg\{-I +\left( 3-\frac{k_0^2}{k^2} \right)\left(1-k^0 L(K) \right) \bigg\} \,, \\
(\Pi^R_{\rm Pow})^{(1)}_{\mathrm{L}} & = \frac{\che^2}{4\pi^2} \frac{K^2}{3} \bigg\{ 1-\frac{\pi^2}{4}+\frac{I^2}{2} -\left(3-\frac{k_0^2}{k^2}\right) k^0 H(K)  \\
  &\hphantom{{}= \frac{\che^2}{4\pi^2} \frac{K^2}{3} \bigg\{ 1}
  -\left[2I-(3-I)\left(1-\frac{k_0^2}{k^2}\right)\right]\left(1-k^0 L(K)\right) \bigg\} \,.
\end{split}    
\end{equation}
Here, the UV divergence multiplying $(\mu/\overline{\Lambda})^{-2\varepsilon}$ in the integration measure has resulted in explicit factors of $\log(\mu/\overline{\Lambda})$ that have been absorbed in the function
\begin{equation}
 I \equiv 2-2\log\left(\frac{2\mu}{\overline{\Lambda}}\right) \,.
\end{equation}

\section{Discussion}
\label{sec:discussion}
\noindent
In the paper at hand, we  determined the NLO self-energy of soft photons traversing a hot and dense electromagnetic plasma. Our computation generalizes the results obtained in \Refs\cite{Manuel:2016wqs,Carignano:2017ovz,Carignano:2019ofj} to nonzero electron chemical potential, and paves the way for a future similar calculation in the context of QCD. Our main motivation for this work stems from a desire to extend the determination of the pressures of cold and dense QED and QCD to full N$^3$LO, where NLO self-energies are required for the proper physical dressing of photon (or gluon) propagators. Indeed, the present article is the long companion paper of a letter~\cite{Seppanen}, where we determine the $O(\alpha_e^3)$ pressure of dense zero-temperature QED up to one undetermined coefficient stemming from the hard sector of the theory. For this reason, we also determined a number of $O(\epsilon)$ parts to the LO and NLO self-energies (see \Sec\ref{coldstuff}) which are needed in the pressure calculation.

Our main result for the NLO contribution to the photon self-energy can be found from \Sec\ref{subsec:resultssum}. Similarly to the zero-$\mu$ case \cite{Carignano:2019ofj}, we find that the UV and IR singularities are fully canceled between the different dimensionally regularized two-loop self-energy diagrams, and that the resulting final expression for the NLO contribution is finite for $D=4$. We also find that, at non-vanishing $T$ and $\mu$, the NLO contribution in \eq\eqref{eq:finalNLOresults} contains a very nontrivial medium dependence, where the medium-dependent mass scale $T^2/3 + \mu^2/\pi^2$ does not factorize out as in the LO case in \eq\eqref{eq:piLOmain}. This is due to the more complicated radial integral structures present in the two-loop diagrams. Interestingly, our result for the NLO longitudinal self-energy introduces a new HTL structure with a squared logarithm [see \eq\nr{eq:finalNLOresults}]. This term is solely generated by non-vanishing $\mu$. 

As a physical application of our result, in \Sec\ref{subsec:photonmodes} we studied the transverse and longitudinal components of the soft photon propagator at NLO. Specifically, we computed the plasmon dispersion relations and the residues of the corresponding poles in the time-like region, along with the spectral function in the Landau cut in the space-like region. The plasma frequency and Debye mass were calculated as specific limits of these results, while a contribution to the asymptotic mass was also obtained. We found that the NLO HTL-resummed propagator is well-behaved in the region of soft momenta, $k \lesssim \mE$, except in the vicinity of the light cone, where the HTL-resummation breaks down. In this region, we suspect that further resummations are necessary.

Finally, it is worth acknowledging that while this article deals with QED and photon propagation therein, to a large extent this calculation represents a simpler test case for a similar forthcoming computation in QCD. In QED, the absence of gauge field self-interactions leads to an overall lower number of two-loop diagrams contributing to the gauge-boson self-energy at NLO, while also simplifying the soft limit of the self-energy at LO. Likewise, QED obeys simple Ward identities, reducing the number of basis tensors with potentially nonzero coefficients (see e.g.~\Ref\cite{Weldon:1996kb}). However, the methods developed and utilized in this work should mostly suffice for QCD as well. This computation is  already underway.

\tocless{\section*{Acknowledgements}}
We thank Peter Arnold and Jacopo Ghiglieri for useful discussions. JÖ, RP, PS, KS, and AV have been supported by the Academy of Finland grant no.~1322507, as well as by the European Research Council, grant no.~725369. TG was supported in part by the Deutsche Forschungsgemeinschaft
(DFG, German Research Foundation) -- Project-ID 279384907 -- SFB 1245
and by the State of Hesse within the Research Cluster ELEMENTS 
(Project ID 500/10.006). In addition, KS gratefully acknowledges support from the Finnish Cultural Foundation. JÖ acknowledges financial support from the Vilho, Yrjö and Kalle Väisälä Foundation of the Finnish Academy of Science and Letters.

\newpage
\appendix

\section{Radial and angular integrals}
\label{app:integrals}

\subsection{Radial integrals}
Here we list some results for radial integrals of the distribution functions defined in \eq\eqref{eq:distfuncs}. To save space, we denote the radial part of the integration measure in $d=3-2\epsilon$ spatial dimensions (see \eq\eqref{eq:intmeasure}) by
\begin{equation}
\int_p \equiv \int_0^\infty \mathrm{d}p \, p^{d-1} \,.
\end{equation}
The results for integrals over the bosonic and fermionic distribution functions may be written in terms of the polylogarithm function as
\begin{align}
  \int_p p^{\alpha} N_B^{ }(p) &= T^{d+\alpha} \Gamma(d+\alpha) \mathrm{Li}_{d+\alpha}\left(1\right) \,, \label{eq:rintbos} \\
\int_p p^{\alpha} N_F^\pm(p) &= T^{d+\alpha} \Gamma(d+\alpha) \mathrm{Li}_{d+\alpha}\bigl(-\exe^{\mp\frac{\mu}{T}}\bigr) \,,
\end{align}
where $\alpha$ is a parameter and scale-free parts of the integrands have been discarded as they vanish in dimensional regularization. Upon integration by parts, we obtain results for the derivatives of the distribution functions,
\begin{align}
  \int_p p^\alpha \frac{\mathrm{d}}{\mathrm{d}p} N_B^{ }(p) &= - T^{d+\alpha-1} \Gamma(d+\alpha) \mathrm{Li}_{d+\alpha-1}\left(1\right) \,, \\
\int_p p^\alpha \frac{\mathrm{d}}{\mathrm{d}p} N_F^\pm(p) &= - T^{d+\alpha-1} \Gamma(d+\alpha) \mathrm{Li}_{d+\alpha-1}\bigl(-\exe^{\mp\frac{\mu}{T}}\bigr) \label{eq:rintderferm}  \,.
\end{align}

For the self-energy calculations, we need the leading terms of the small $\epsilon$ expansions for various combinations of the above integrals for particular values of $\alpha$,
\begin{align}
  \mathcal{R}_1 &\equiv \int_p \frac{1}{p} N_B^{ }(p) = \frac{\pi^2 T^2}{6} + O(\epsilon) \,, \label{eq:r1int} \\
\mathcal{R}_2 &\equiv \int_p \frac{1}{p} \left(N_F^-(p)+N_F^+(p)\right) = -\frac{\pi^2 T^2+3\mu^2}{6} + O(\epsilon) \,, \\
\mathcal{R}_3 &\equiv \int_p \frac{1}{p^3} \left(N_F^-(p)+N_F^+(p)\right) = \frac{1}{2\epsilon} - \log (\exe^{-\gamE} T) + \mathrm{Li}_0^{(1)}(\exe^{\frac{\mu}{T}}) + \mathrm{Li}_0^{(1)}(\exe^{-\frac{\mu}{T}}) + O(\epsilon) \,, \\
\mathcal{R}_4 &\equiv \int_p \frac{1}{p^2} \left(N_F^-(p)-N_F^+(p)\right) = -\mu + O(\epsilon) \\
\mathcal{R}_5 &\equiv \int_p \frac{1}{p^2} \frac{\mathrm{d}}{\mathrm{d}p} \left(N_F^-(p)+N_F^+(p)\right) = 1 + O(\epsilon) \label{eq:r5int} \,,
\end{align}
where we have used the notation $\mathrm{Li}^{(1)}_0(z) \equiv \lim_{s \to 0}\frac{\partial \mathrm{Li}_s(z)}{\partial s}$. 

The zero-temperature limit of the radial integrals in \eqs\eqref{eq:rintbos}--\eqref{eq:rintderferm} is straightforward to obtain given the following limiting behavior of the polylogarithm
\begin{align}
&\lim_{x\to\infty} \mathrm{Li}_s (-\exe^{-x}) = 0 \,, \\
&\lim_{x\to\infty} \mathrm{Li}_s (-\exe^{x}) = -\frac{x^s}{\Gamma(s+1)} \,, \quad s \not\in \mathbb{Z}^- \,,
\end{align}
where $x=\mu/T$. In this limit, we can compactly write the $d$-dimensional results for the integrals defined in \eqs\eqref{eq:r1int}--\eqref{eq:r5int} as
\begin{align}
\mathcal{R}_1 &= 0 \,, \\
\mathcal{R}_2 &= -\frac{\mu^{d-1}}{d-1} \,, \\
\mathcal{R}_3 &= -\frac{\mu^{d-3}}{d-3} \,, \\
\mathcal{R}_4 &= -\frac{\mu^{d-2}}{d-2} \,, \\
\mathcal{R}_5 &= \mu^{d-3} \,.
\end{align}

\subsection{Angular integrals}
In $d=3-2\epsilon$ spatial dimensions, the angular part of the integration measure in \eq\eqref{eq:intmeasure} may be written as
\begin{equation}
\int_z \equiv \int_{-1}^1 \mathrm{d}z \, (1-z^2)^\frac{d-3}{2} \,,
\end{equation}
where $z=\hat{\mathbf{k}}\cdot\hat{\mathbf{p}}$ parametrizes an angle between an external spatial unit vector $\hat{\mathbf{k}}$ and the spatial unit vector in the direction of the loop momentum $\hat{\mathbf{p}}$. According to our conventions, the “direction” of the on-shell loop four-momentum $\Pc$ is denoted by $v = (1,\hat{\mathbf{p}})$, so that $v \cdot \Kc = -k^0+k z$ for the external four-momentum $\Kc$. In the HTL limit, we often encounter integrals of the type
\begin{equation}
\label{eq:Aalphaint}
\mathcal{A}_\alpha \equiv \int_z \frac{1}{(v\cdot\Kc)^\alpha} = \frac{\Gamma\left(\frac{1}{2}\right)\Gamma\left[\frac{1}{2}(d-1)\right]}{\Gamma\left(\frac{d}{2}\right)} (-k^0)^{-\alpha} {}_2F_1\left(\frac{\alpha}{2},\frac{1+\alpha}{2};\frac{d}{2};\frac{k^2}{k_0^2}\right) 
\end{equation}
for some parameter $\alpha$. The result has been written in terms of the hypergeometric function ${}_2F_1$. The following recursion relation applies for $\alpha \in \mathbb{Z}^+$,
\begin{equation}
\mathcal{A}_{\alpha+1} = \frac{1}{\alpha} \frac{\mathrm{d}}{\mathrm{d}k^0} \mathcal{A}_\alpha \,.
\end{equation}

Another useful integral is given by
\begin{equation}
\label{eq:Aialphaint}
\mathcal{A}_\alpha^i \equiv \int_z \frac{v^i}{(v\cdot\Kc)^\alpha} = \hat{k}^i \int_z \frac{z}{(v\cdot\Kc)^\alpha} = \frac{k^i}{k^2} \big(\mathcal{A}_{\alpha-1}+k^0 \mathcal{A}_\alpha\big) \,,
\end{equation}
where we have exploited  spatial rotational symmetry. By denoting 
\begin{equation}
L(\Kc) \equiv \frac{1}{2k}\log\frac{k^0+k}{k^0-k} 
\;,
\end{equation}
we write the small $\epsilon$ expansions of the above results at specific values of $\alpha$ as
{
\allowdisplaybreaks
\begin{align}
\mathcal{A}_0 &= \int_z 1 = 2 + (1-\log(2))4\epsilon + O(\epsilon^2) \,, \\
\begin{split}
\mathcal{A}_1 &= \int_z \frac{1}{v\cdot\Kc} = -2 L(\Kc) + \bigg\{ 2L(\Kc) \log\left(\frac{\Kc^2}{k^2}\right) \\
  &\hphantom{{}= \int_z \frac{1}{v\cdot\Kc} = -2 L(\Kc) }
  +\frac{1}{k}\left[ \mathrm{Li}_2\left(\frac{k^0-k}{k^0+k}\right) - \mathrm{Li}_2\left(\frac{k^0+k}{k^0-k}\right) \right] \bigg\}\epsilon + O(\epsilon^2) \,,
\end{split} \\
\mathcal{A}_2 &= \int_z \frac{1}{(v\cdot\Kc)^2} = -\frac{2}{\Kc^2} +\frac{4}{\Kc^2}\left(\log(2)-k^0 L(\Kc)\right) \epsilon + O(\epsilon^2) \,, \\
\begin{split}
\mathcal{A}_3 &= \int_z \frac{1}{(v\cdot\Kc)^3} = -\frac{2k^0}{(\Kc^2)^2} -\frac{2}{(\Kc^2)^2}\big\{(1-2\log(2))k^0 \\
  &\hphantom{{}= \int_z \frac{1}{(v\cdot\Kc)^3} = -\frac{2k^0}{(\Kc^2)^2}}
  + \left(k_0^2+k^2\right)L(\Kc)\big\} \epsilon + O(\epsilon^2) \,,
\end{split} \\
\begin{split}
\mathcal{A}_4 &= \int_z \frac{1}{(v\cdot\Kc)^4} = -\frac{2}{3}\frac{3k_0^2+k^2}{(\Kc^2)^3} -\frac{2}{3}\frac{2}{(\Kc^2)^3}\big\{(2-3\log(2)+k^0 L(\Kc))k_0^2 \\ 
  &\hphantom{{}= \int_z \frac{1}{(v\cdot\Kc)^4} = -\frac{2}{3}\frac{3k_0^2+k^2}{(\Kc^2)^3}}
  + (1-\log(2)+3k^0 L(\Kc))k^2 \big\} \epsilon + O(\epsilon^2) \,.
\end{split}
\end{align}
}

\bibliography{references}

\end{document}